\documentclass[floatfix,secnumarabic,amssymb,preprint,nobibnotes,nofootinbib,aps,pra,showpacs,superscriptaddress]{revtex4-2}
\usepackage{bbold}
\usepackage{siunitx}
\usepackage{comment}
\usepackage{mathtools}
\usepackage{amssymb}
\usepackage{amsmath}
\usepackage{mathrsfs}
\usepackage{graphicx}
\usepackage{svg}
\usepackage{epsfig}
\usepackage{dcolumn}
\usepackage{color}
\usepackage{xcolor}
\usepackage{bm}
\usepackage{braket}
\usepackage{soul}
\usepackage{subfigure}
\usepackage{appendix}
\usepackage{float}
\usepackage{times}
\usepackage[colorlinks=true, citecolor=red, linkcolor=blue, urlcolor=blue]{hyperref}
\usepackage{cleveref}
\usepackage{setspace}
\newcommand{\be}{\begin{equation}}
\newcommand{\ee}{\end{equation}}
\newcommand{\bea}{\begin{eqnarray}}
\newcommand{\eea}{\end{eqnarray}}
\newcommand{\ba}[1]{\begin{array}{#1}}
\newcommand{\ea}{\end{array}}

\begin{document}

\title{\textbf{{\large \bfseries {{ Quantum synchronization in atom-cavity coupled systems}}}}\\[0.4cm]}
\author{Katha Haldar}
\affiliation{School of Physical Sciences, Indian Association for the Cultivation of Science, Jadavpur, Kolkata 700032, India.}
\author{ Anushree Dey}
\affiliation{School of Physical Sciences, Indian Association for the Cultivation of Science, Jadavpur, Kolkata 700032, India.}
\author{Saikat Ghosh}
\affiliation{Department of Physics, Indian Institute of Technology, Kanpur, India}
\author{Bimalendu Deb}
\email{msbd@iacs.res.in}
\affiliation{School of Physical Sciences, Indian Association for the Cultivation of Science, Jadavpur, Kolkata 700032, India.}
\affiliation{CQuERE, TCG Centres of Research and Education in Science and Technology, Salt Lake, Kolkata, India.}
\begin{abstract}
We consider a  cavity quantum electrodynamics (CQED) setup where an ensemble of spin-1 atoms driven by a classical field interacts with two driven cavity field modes. We show that the two sub-systems - the atoms and the two-mode cavity field,  exhibit limit-cycle or quantum synchronization depending on  the cavity drive strengths and other system parameters. In order to visualize the limit cycle and synchronization in the field, we introduce generalized two-mode field Husimi-Q function using SU(2) coherent states. For atomic spin states, we make use of spin-1 Husimi-Q function to visualize the same. We further corroborate the occurrence of field synchronization and limit cycle by analyzing $U(1)$ symmetry breaking, Wigner quasi-probability distribution, coherence  properties of reduced field density matrix. Our results reveal intriguing interplay between the limit-cycle and quantum synchronization behavior,  highlighting emergent quantum phenomena in driven dissipative interacting quantum systems. Additionally, we calculate Wehrl mutual information between the two modes and find that the field synchronization and Wehrl mutual information are mutually exclusive. Our study would be important to extract quantum advantages in quantum metrology using coupled atom-cavity systems.
\end{abstract}
\maketitle
\section{Introduction}
Synchronization  is one of the most striking collective phenomena in nonlinear science. It happens when multiple dynamical systems adjust their motion to match a rhythm through  mutual interactions. A weak perturbation to  self-sustained oscillators may lead to synchronization \cite{Lee2013PRA,Pikovsky2003}. A self-sustained oscillator is an autonomous active system containing an internal source of energy that may drive the system into periodic motion.  For a classical noiseless system, this motion can be represented by a \textit{limit cycle} which is a closed curve in phase-space \cite{vanDerPol1926,vanDerPolMark1928}. Existence of a limit cycle may be a prerequisite for achieving synchronization through external perturbation. 

In contrast to classical synchronization, quantum synchronization is a relatively new concept that has attracted a lot of research interest in recent times \cite{Mari2013,Giorgi2012}.  In the quantum regime, however, phase diffusion, measurement back-action and finite Hilbert-space make synchronization fundamentally different from its classical counterpart \cite{Roulet2018spin, Roulet2018ent, vanDerPol1926, Koch2011}. In open quantum systems, the relative phase between quantum oscillators is hindered by dephasing and damping due to quantum fluctuations \cite{Breuer2002,Rivas2012}.  As a result, the relative phase can not attain a fixed value, but it may settle down  to a certain average value within a narrow range in the limit cycle leading  to the phenomenon called phase-locking \cite{Lorch2016,Lorch2017}. Phase-space quasi-probability distribution widely used in quantum optics plays an important role in describing quantum synchronization \cite{Husimi1940,Schleich2001}.  A quantitative measure of quantum synchronization in continuous-variable systems was introduced by Mari {\it{et al.}} \cite{Mari2013}. Hush {\it et al.} \cite{PhysRevA.91.061401} proposed a measure of quantum synchronization in terms of the maximum difference between the phase-space distribution in the synchronized state  and that in the limit-cycle state. 

In the context of quantum synchronization, a particularly interesting research direction is the search for experimentally accessible signatures of quantum limit cycles and phase locking. Tomographic and photon-correlation diagnostics have been used to characterize synchronization in driven van der Pol oscillators \cite{Adhikary_2026}, In such diagnostic methods, continuous heterodyne monitoring provides a measurement-based picture that can relate the limit cycle and synchronization to observables available in output fields \cite{c5wj-4s73}. Spin-based and few-level systems such as the quantum spin van der Pol oscillator and twin limit-cycle oscillators have been employed to examine a variety of measurable quantities of quantum-synchronized state. Recent experimental progress in trapped-ion van der Pol oscillators~\cite{liu2025observationsynchronizationquantumvan} and seeded Floquet-Rydberg systems~\cite{liu2025quantumsynchronizationcriticalpoint} further gives an impetus to the field of quantum synchronization and its applications. More recently, quantum synchronization has also been connected to dissipative sensing through quantum Fisher information \cite{Vaidya2025}, highlighting the metrological usefulness of synchronized quantum dynamics.  These developments motivate us to address the question whether atom-mediated coupling between two driven cavity modes can induce a two-mode field quantum limit cycle and generate phase-locked field dynamics.

In this work, we explore the synchronization of field dynamics in a CQED platform \cite{Haroche2006,Raimond2001,Walther2006}. We consider an ensemble of ground-state hyperfine spin $F=1$ atoms interacting with a bimodal cavity field \cite{StamperUeda2013,Barberena2019}. Our model is schematically shown in Fig.~1. The three magnetic sublevels $m_F = 0, \pm 1$ are split due to a magnetic field applied along the cavity axis. The level $m_F = 0$ is coupled to a $P$-excited state with hyperfine spin $F'$, $m_{F'} = 0$ by a $\pi$-polarized laser field. The excited state is assumed to be significantly detuned from all other detuning and damping parameters, allowing it to be adiabatically eliminated \cite{Brion2007},  rendering the atom into an effective spin-1 system. The two cavity modes with frequencies $\omega_1$ and $\omega_2$ are left and right circularly polarized and tuned near transitions $m_F=-1 \leftrightarrow m_{F'} = 0$ and $m_F = 1 \leftrightarrow m_{F'} = 0$, respectively. Both the cavity modes are driven by two laser fields.

We show that the coupled atom-field dynamics evolve into limit cycles, enabling us to identify  quantum synchronization in the two-mode cavity field \cite{Koppenhoefer2019,Koppenhoefer2020,Nigg2018} as well as in the atomic system due to appropriate drive and dissipation in the system. To analyze the system in terms of Husimi-Q quasi-probability distribution functions \cite{Husimi1940,Glauber1963}, we employ SU(2) coherent states for the fields using Schwinger's angular momentum representation of bosons  and spin-coherent states for the atoms \cite{Radcliffe1971,Arecchi1972,Agarwal1981,ZhangFeng1990}. Schwinger's representation of bosons in terms of angular momentum is generally applicable in a closed system where the total number of bosons is conserved. However, since we are dealing with an open quantum system, we develop an operational  approach to the application of SU(2) coherent states to the two-mode cavity field: We sub-divide the Fock space of two-mode field into a large number (say, $N_{\rm max}$) of total photon number preserving sectors and consider SU(2) coherent states of each sector and then make a coherent superposition of all the SU(2) field coherent states where the superposition coefficients are determined by the actual density matrix. We finally allow  $N_{\rm max}$ to go to infinity and thereby relax the photon number conserving constraint. This procedure is akin to what is used in quantum phase operator formalism.\cite{PeggBarnett1988UnitaryPhase}. We  solve atom-field joint density matrix by quantum trajectory method
Our findings are illustrated not only in terms of the Husimi-Q functions, but also by looking at  characteristic  features in the reduced field and atomic density matrices. We find that, when there is no cavity driving and only atoms are driven by a classical field, both the  field and the atoms exhibit limit cycles but no synchronization. On the other hand, if both the modes are driven by two weak classical fields, the limit cycle states of the field and atomic sub-systems are shown to evolve into out-of-phase synchronized states. The phase locking in synchronized states is shown to be due to exsistence of the off-diagonal density matrix elements or phase coherences. It is important to note that we achieve quantum synchronization only in the strong-coupling cavity QED regime where the two cavity decay constants are much smaller than the atom-field coupling constants and the cavity drive strengths, implying that coherent processes must dominate over the incoherent ones for the quantum synchronization. To explore the quantum information aspects of synchronized states, we calculate the Wehrl mutual information \cite{PhysRevA.103.062222}. Our results show that the Wehrl mutual information is minimum when the synchronization is maximum and vice versa. Overall, this study offers physical insight into synchronized quantum dynamics of coupled atom-cavity systems which may find potential applications in quantum metrology \cite{Degen2017,Giovannetti2011,Pezze2018,Braun2018}.

The paper is organized as follows. In Sec.\ref{Model}, we introduce the model and outline the formulation of the problem. Here we describe the method used to characterize the synchronization of the bimodal cavity field and atom. In Sec.~\ref{Discussion}, we present and discuss our main results. Finally, Sec.~\ref{Conclusions} provides concluding remarks.

\section{ The model: Formulation of the problem}\label{Model}
Our model is schematically shown in Fig.1. We consider an ensemble of atoms initially prepared in ground-state hyperfine spin $F = 1$ with $m_F = 0$ inside a two-mode cavity by optical pumping. A biased magnetic field $B$ is applied along the cavity axis to lift the degeneracy of  the three magnetic  sub-levels. The three ground-state sub-levels are denoted as $\mid a \rangle = \mid F=1, m_F = - 1 \rangle$, $\mid b \rangle = \mid F=1, m_F = 0 \rangle$ and $\mid c \rangle = \mid F=1, m_F = 1 \rangle$. The states $\mid a \rangle$ and $\mid c \rangle $ are shifted from $\mid b \rangle$ by 
+$\hbar \Delta_B$ and $ -\hbar \Delta_B$, respectively. The state $\mid b \rangle $ is coupled to an excited state $\mid d \rangle = \mid F'=2, m_{F'} = 0\rangle $ by a $\pi $-polarized laser of frequency $\omega_{l}$ while the states $\mid a \rangle $  and $\mid c \rangle $ are coupled to $\mid d \rangle $ by left and right circularly polarized cavity modes of frequency $\omega_1$ and $\omega_2$, respectively. 

\begin{figure}[H]
\centering
\includegraphics[width=1.2\linewidth]{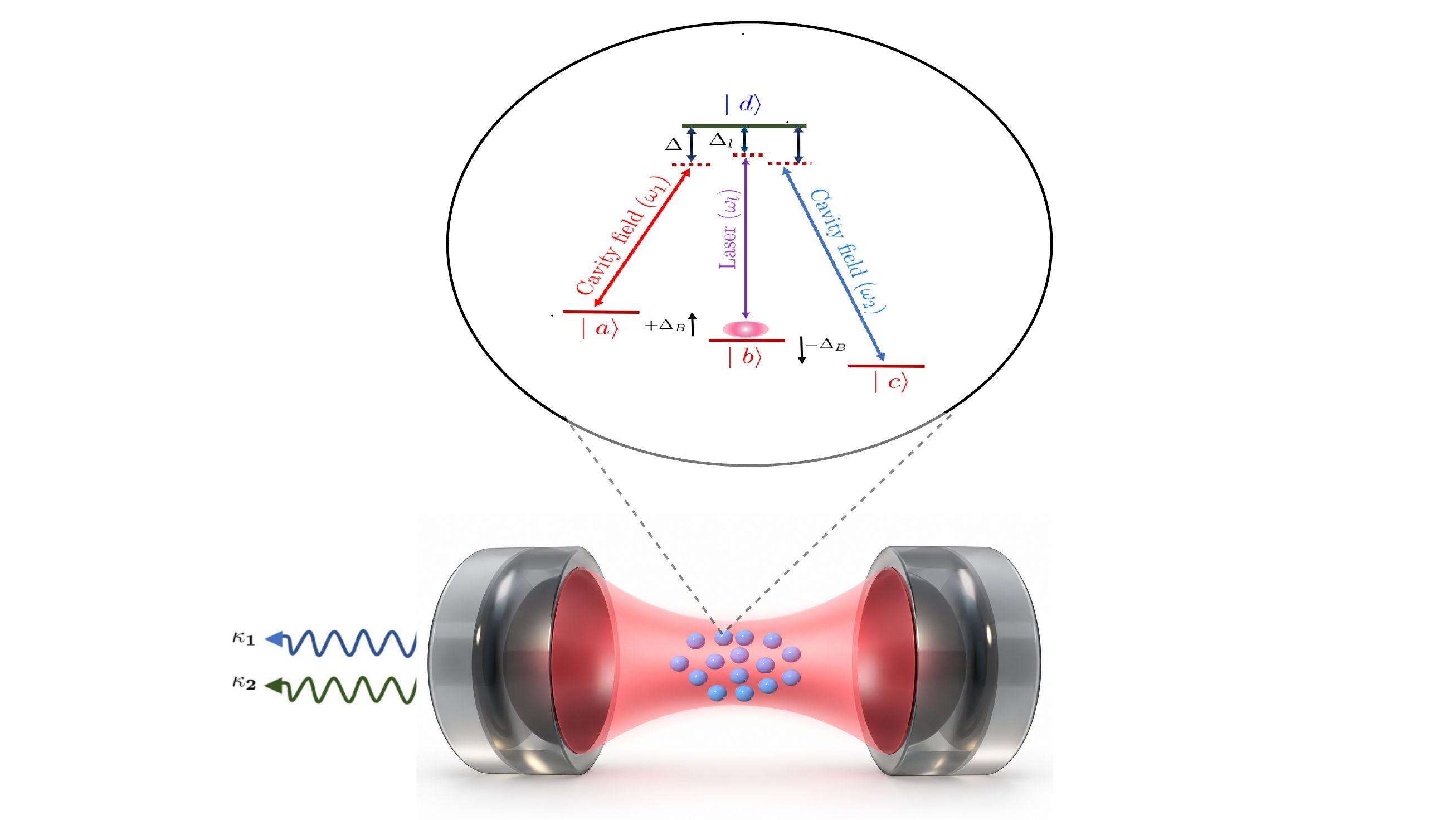}\hspace*{-0.2cm}
\caption{A schematic diagram showing an ensemble of atoms in a bimodal cavity, $\kappa_1$ and $\kappa_2$ are the two cavity decay rates. Inset shows a single four-level atom interacting with two cavity modes and a laser field. The levels  $\mid a \rangle$, $\mid b \rangle$ and $\mid c \rangle$ denote three ground-state magnetic sub-levels which are split due to an applied magnetic field. The cavity modes 1 and 2 couple the  levels $\mid a \rangle $ and $\mid c \rangle $ to the excited level 
$\mid d \rangle$, respectively; while the transition $\mid b \rangle  \rightarrow \mid d \rangle $ is driven by a laser field.}
\end{figure}

\subsection{The Hamiltonian and the master equation}
\vspace{-0.5 cm}
The  Hamiltonian of the system is
$H = H_{0} + H_{{\rm atom-cavity}} + H_{{\rm drive}}$ where 
\bea 
H_0
= \hbar \displaystyle\sum_{\alpha=a,b,c,d} \omega_\alpha \lvert \alpha \rangle \langle \alpha \rvert
+ \hbar \displaystyle\sum_{i=1}^{2} \omega_i \, \hat{a}_i^\dagger\hat{a}_i
\eea 
represents the free part of the Hamiltonian. Here $\hbar \omega_\alpha$ denotes the  energy  of the atomic level $\lvert \alpha\rangle$ with $\alpha=(a,b,c,d)$,  $\hat{a}_i$ is the photon annihilation operator of the $i$th cavity mode of frequency $\omega_i$. Note that in the absence of a biased magnetic field $\omega_a = \omega_b = \omega_c$ while in the presence of the magnetic field $\omega_a = - \omega_c = \Delta_B$ and $\omega_b = 0$. 
The interaction between the cavity fields and the atoms is given by 
 \bea 
 H_{{\rm atom-cavity}} = \hbar \left [ g_1 \hat{a}_1 \mid d \rangle \langle a \mid + g_2 \hat{a}_2 \mid d \rangle \langle c \mid + {\rm h. c.} \right ] 
 \eea 
 where $g_i$ is the coupling constant between the $i$th cavity mode and the atom. The driving part of the Hamiltonian is 
 \bea 
 H_{{\rm drive}} =  \frac{1}{2} \hbar \left [ \Omega e^{-i \omega_l t} \ket{d}\bra{b} + \sum_{i=1}^{2}  \Omega_i e^{i\omega_{c_{i}}t} \hat{a}_i  + {\rm h.c.} \right ]
 \eea 
The atom is driven by a laser field of frequency $\omega_l$ with  $\Omega$ being the corresponding field-atom Rabi frequency and $\Omega_i$ is the Rabi frequency for the $i$th  cavity drive \cite{deFreitas2017, PhysRevA.48.1400} with $\omega_{c_i}$ being the drive frequency. Here we keep $\Omega_1=\Omega_2=\varepsilon$. We make use of the electric-dipole and rotating wave approximations. Applying the unitary transformation 
\begin{align}
    U(t) = \exp-it[\{({\omega}_{l}-\omega_{c_{1}})\ket{a}\bra{a} + ({\omega}_{l}-\omega_{c_{2}})\ket{c}\bra{c}&+&\nonumber\\ {\omega}_{l}\ket{d}\bra{d}+\omega_{c_{1}}\hat{a}_1^{\dag}\hat{a}_1+\omega_{c_{2}}\hat{a}_2^{\dag}\hat{a}_2\}]
\end{align}
we go to a rotating frame in which the Hamiltonian reads as 
 \begin{eqnarray}
  \Tilde{H} &= &\sum_{\alpha=a,b,c,d} \hbar\Delta_{\alpha} \ket{\alpha}\bra{\alpha} +
 \sum_{i=1,2}\hbar\delta_{i}{\hat{a}_i}^{\dag}\hat{a}_i  \nonumber \\
&+& \hbar{g_1}\hat{a}_1\ket{d}\bra{a} + \hbar{g_2}\hat{a}_2\ket{d}\bra{c}\nonumber\\ 
&+& \nonumber \frac{1}{2} \hbar\Omega\ket{d}\bra{b}+h.c.+ \frac{1}{2}\hbar \sum_{i=1}^{2} \Omega_i \left( \hat{a}_i + \hat{a}_i^{\dagger} \right)
 \end{eqnarray}
 where, $\Delta_a=\omega_a-(\omega_l-\omega_{c_{1}})$, $\Delta_c=\omega_c-(\omega_l-\omega_{c_{2}})$, $\Delta_d=\omega_d-\omega_l$.  The detuning of the two cavity fields are  $\delta_1=\omega_1-\omega_{c_{1}}$\ and, $\delta_2=\omega_2-\omega_{c_{2}}$. Considering that $|\Delta_{d}|$ to be quite large compared to all
other frequency scales in the system, we adiabatically eliminate the level $|d\rangle$ to obtain an effective  Hamiltonian $H^{{\rm ad}}=H_0^{{\rm ad}}+ H_{{\rm int}}^{{\rm ad}}$, 
where
\begin{eqnarray}
H_0^{{\rm ad}} &=&
\hbar \Delta_a \lvert a\rangle\langle a\rvert
+ \hbar \Delta_b \lvert b\rangle\langle b\rvert
+ \hbar \Delta_c \lvert c\rangle\langle c\rvert\nonumber\\
&+& \hbar \delta_1\, \hat{a}_1^{\dagger} \hat{a}_1
+ \hbar \delta_2\, \hat{a}_2^{\dagger} \hat{a}_2
+ H_{{\rm Stark}}
\label{eq:H0}
\end{eqnarray}
\begin{eqnarray}
H_{\rm{Stark}} &=&
- \hbar |g_1|^{2}
\, {\rm Re}\!\left(\frac{1}{D_a}\right)
\, \hat{a}_1^{\dagger} \hat{a}_1 \,
\lvert a\rangle\langle a\rvert\nonumber\\
&-& \hbar \frac{|\Omega|^{2}}{4}
\, {\rm Re}\!\left(\frac{1}{D_b}\right)
\, \lvert b\rangle\langle b\rvert\nonumber\\
&-& \hbar |g_2|^{2}
\, {\rm Re}\!\left(\frac{1}{D_c}\right)
\, \hat{a}_2^{\dagger}\hat{a}_2 \,
\lvert c\rangle\langle c\rvert .
\label{eq:Hstark}
\end{eqnarray}

The interaction terms of the adiabatic Hamiltonian are
\begin{equation}
\small
\begin{aligned}
H_{\mathrm{int}}^{\mathrm{ad}}
={}&
-\frac{\hbar g_1 \Omega^{*}}{4}
\left(
\frac{1}{D_a}
+
\frac{1}{D_b^{*}}
\right)
\hat{a}_1 |b\rangle\langle a|
-
\frac{\hbar g_2 \Omega^{*}}{4}
\left(
\frac{1}{D_c}
+
\frac{1}{D_b^{*}}
\right)
\hat{a}_2 |b\rangle\langle c|
\\
&-
\frac{\hbar g_1 g_2^{*}}{2}
\left(
\frac{1}{D_a}
+
\frac{1}{D_c^{*}}
\right)
\hat{a}_2^{\dagger}\hat{a}_1 |c\rangle\langle a|
+
\mathrm{h.c.}+\sum_{i=1,2}
\frac{\hbar \Omega_i}{2}
(\hat{a}_i+\hat{a}_i^{\dagger})
\end{aligned}
\label{eq:Heffad}
\end{equation}
where 
\begin{eqnarray}
D_a &=&
\Delta_d - \Delta_a - \delta_1
- \frac{i\Gamma_d}{2},
\qquad\nonumber\\
D_b &=&
\Delta_d - \Delta_b
- \frac{i\Gamma_d}{2},
\qquad \nonumber\\
D_c &=&
\Delta_d - \Delta_c - \delta_2
- \frac{i\Gamma_d}{2}.
\label{eq:complexdet}
\end{eqnarray}
are the complex detunings. 

We next construct master equation for joint atom-field  density matrix  
$\rho_{af}$ in Lindblad form 
\begin{eqnarray}
\frac{\partial \rho_{af}}{\partial t}
&=&
-\frac{i}{\hbar}
\left[
H_{\rm adi},\rho_{af}
\right]
+
\sum_{i=1,2}
\mathcal{D}\!\left[L_{\kappa_i}\right]\rho_{af}
+
\sum_{j=1,2}
\mathcal{D}\!\left[L_{\gamma_j}\right]\rho_{af} \nonumber\\
&+&
\sum_{\alpha=a,b,c}
\mathcal{D}\!\left[L_{{\rm eff},d\alpha}\right]\rho_{af}.
\label{eq:master}
\end{eqnarray}
where the dissipator is defined as
\begin{align}
\mathcal{D}[L]\rho_{af}
=
L\rho_{af}L^\dagger
-
\frac{1}{2}
\left\{
L^\dagger L,\rho_{af}
\right\}.
\label{eq:dissipator}
\end{align}
\begin{align}
L_{\kappa_i}
&=
\sqrt{\kappa_i}\,\hat{a}_i,
\qquad i=1,2,
\\
L_{\gamma_1}
&=
\sqrt{\gamma_1}\,\lvert b\rangle\langle a\rvert,
\qquad
L_{\gamma_2}
=
\sqrt{\gamma_2}\,\lvert b\rangle\langle c\rvert .
\end{align}
The effective jump operators are 
\begin{eqnarray}
L_{{\rm eff},d\alpha}
&=&
\sqrt{\Gamma_{d\alpha}}
\left[
\frac{g_1 \hat{a}_1}{D_a}
|\alpha\rangle\langle a|
+
\frac{\Omega/2}{D_b}
|\alpha\rangle\langle b|
+
\frac{g_2 \hat{a}_2}{D_c}
|\alpha\rangle\langle c|
\right], \nonumber\\
&&
\alpha=a,b,c .
\label{eq:Leff}
\end{eqnarray}

\subsection{Generalized Field Husimi-$Q$ Function}

To discuss the limit cycle and quantum synchronization dynamics in a two-mode quantized field, we make use of the Schwinger spin or angular momentum representation of two-mode bosons \cite{dubus2024schwinger}. If the total number of bosons in the two modes is conserved, then the bosonic operators $\hat{a}_i$ ($i=1,2$) can be expressed by the following angular momentum-like operators 
\begin{align}
{\bf J}_1 &= \tfrac{1}{2}\!\left(\hat{a}_1^\dagger \hat{a}_2 + \hat{a}_2^\dagger \hat{a}_1\right), \\
{\bf J}_2 &= \tfrac{1}{2i}\!\left(\hat{a}_1^\dagger \hat{a}_2 - \hat{a}_2^\dagger \hat{a}_1\right), \\
{\bf J}_3 &= \tfrac{1}{2}\!\left(\hat{a}_1^\dagger \hat{a}_1 - \hat{a}_2^\dagger \hat{a}_2\right),
\end{align}
which satisfy the $SU(2)$ algebra
$[{\bf J}_a,{\bf J}_b] = i\epsilon_{abc} {\bf J}_c$, with $\epsilon_{abc}$ the
Levi--Civita tensor. The total excitation operator is
\begin{align}
{\bf J} = \tfrac{1}{2}\!\left(\hat{a}_1^\dagger \hat{a}_1 + \hat{a}_2^\dagger \hat{a}_2\right)
\end{align}
On any subspace of fixed total
photon number,  the Casimir satisfies
${\bf J}^2 = {\bf J}_1^2 + {\bf J}_2^2 + {\bf J}_3^2$.

The simultaneous eigenstates of ${\bf J}$ and ${\bf J}_3$ are $\ket{j,m}$ with
$m=-j,\dots,j$,
\begin{align}
{\bf J}\ket{j,m} = j\ket{j,m}, \qquad {\bf J}_3\ket{j,m} = m\ket{j,m},
\end{align}
and may be generated from the two-mode vacuum as
\begin{align}
\ket{j,m}
= \frac{(\hat{a}_1^\dagger)^{j+m}(\hat{a}_2^\dagger)^{j-m}}
       {\sqrt{(j+m)!\,(j-m)!}}\,\ket{0,0}.
\end{align}

For a fixed total photon number $N=N_1+N_2$ 
, we have  $j = N/2$.  The
$SU(2)$ coherent state in the two-mode Fock basis is given by 
\begin{align}
\ket{\zeta;N}
= (1+|\zeta|^2)^{-N/2}
  \sum_{k=0}^{N}\sqrt{\binom{N}{k}}\,\zeta^{k}\,\ket{k,N-k},
\label{eq:su2coherent}
\end{align}
parameterized by
\begin{align}
\zeta = \tan\!\left(\tfrac{\theta}{2}\right) e^{i\phi},
\qquad
\theta\in[0,\pi],\ \phi\in[-\pi,\pi].
\label{eq:zeta-def}
\end{align}

In the present driven-dissipative system,  the total photon number is
not conserved: coherent driving continuously injects photons while
cavity loss removes them.  So,  the steady-state density matrix
may be considered as a linear combination of $N_{\rm max}$ sub-density matrices corresponding to different $N$-photon sectors where $N_{\rm max} \rightarrow \infty$. We thus introduce first
a shell-projected Husimi-Q distribution function for a two-mode cavity field  restricted to a single
fixed-$N$ and then construct a generalized Husimi-Q function by combining different $N$ sectors as we describe below. 

 Let $\rho_f$
denote the reduced density matrix of the two-mode cavity field in
the Fock basis $\{\ket{n_1,n_2}\}$. For each sector $N$, we  define
$\ket{\zeta;N}$ through Eq.~\eqref{eq:su2coherent} and a
\emph{reference state} on the Bloch sphere by the coherent
superposition
\begin{align}
\ket{\Psi_\alpha(\theta,\phi)}
= \sum_{N=0}^{N_{\max}} \alpha_N\,\ket{\zeta;N},
\qquad
\sum_{N=0}^{N_{\max}} |\alpha_N|^2 = 1.
\label{eq:Psi-alpha}
\end{align}
The generalized field Husimi distribution is then given by 
\begin{align}
Q_{\mathrm{field}}(\theta,\phi)
&\equiv
\bra{\Psi_\alpha(\theta,\phi)}\,\rho_f\,\ket{\Psi_\alpha(\theta,\phi)}
\nonumber\\
&=
\sum_{N,N'=0}^{N_{\max}}
\alpha_N^{*}\,\alpha_{N'}\,
\bra{\zeta;N}\rho_f\ket{\zeta;N'}.
\label{eq:qgen}
\end{align}
The amplitudes $\alpha_N$ fix the relative weight with which the
density matrix is probed in each  sector. After all calculations are done, we let $N_{\rm max}$ go to infinity - a procedure akin to the quantum phase operator formalism \cite{PeggBarnett1988UnitaryPhase}.

In the numerical calculations reported here,  the sector amplitudes
are set by the square root of the steady-state sector populations,
\begin{align}
\alpha_N = \sqrt{p_N},
\qquad
p_N = \mathrm{Tr}\!\left[\rho_f\,P_N\right]
    = \sum_{k=0}^{N}\bra{k,N-k}\rho_f\ket{k,N-k},
\label{eq:alpha-pop}
\end{align}
where $P_N$ is the projector onto the $N_1+N_2=N$ sector, with the normalization
$\sum_N |\alpha_N|^2 = \mathrm{Tr}\,\rho_f = 1$ and weights each
sector by its actual occupation in the steady state.  So,  the
distribution depends not only on sector populations but also on the inter-sector coherences
that can lead to relative-phase locking under appropriate physical conditions.

Substituting Eq.~\eqref{eq:su2coherent} into
Eq.~\eqref{eq:qgen} and using \eqref{eq:zeta-def} we obtain
\begin{align}
Q_{\mathrm{field}}(\theta,\phi)
={}&
\sum_{N,N'=0}^{N_{\max}}
\alpha_N^{*}\,\alpha_{N'}\,
\left[1+\tan^{2}\!\tfrac{\theta}{2}\right]^{-(N+N')/2}
\nonumber\\
&\times
\sum_{k=0}^{N}\sum_{k'=0}^{N'}
\sqrt{\binom{N}{k}\binom{N'}{k'}}
\left[\tan\!\tfrac{\theta}{2}\right]^{k+k'}
e^{\,i(k'-k)\phi}
\nonumber\\
&\times
\bra{k,N-k}\rho_f\ket{k',N'-k'}.
\label{eq:qgen_expanded}
\end{align}
Unlike a shell-projected Husimi function,
Eq.~\eqref{eq:qgen_expanded} retains the off-diagonal matrix elements
$\bra{k,N-k}\rho_f\ket{k',N'-k'}$ with $N\neq N'$ and therefore
explicitly incorporates coherences between distinct excitation manifolds.
This is essential here because the steady-state field generated by
coherent pumping and dissipation is distributed across multiple
photon-number sectors.

A phase-only summary of $Q_{\mathrm{field}}$ is obtained from the
$\theta$-marginal
\begin{align}
M(\phi) = \int_{0}^{\pi}
Q_{\mathrm{field}}(\theta,\phi)\,\sin\theta\,d\theta,
\end{align}
and the synchronization measure
\begin{align}
S_{\mathrm{field}}(\phi)
= M(\phi) - \frac{1}{2\pi}\!\int_{-\pi}^{\pi}M(\phi')\,d\phi'.
\label{eq:Sphi_def}
\end{align}
For a limit-cycle state with unbroken $U(1)$ symmetry,
$M(\phi)$ is independent of $\phi$ and
$S_{\mathrm{field}}(\phi)\equiv 0$. A localized peak in
$S_{\mathrm{field}}(\phi)$ therefore signals breaking of the
relative-phase symmetry between the two cavity modes and the
emergence of phase synchronization.

\subsection{Atomic Husimi Q function}

We  write spin coherent state for spin $S$ by rotating the  state $\ket{S,m_s=S}$ as given by 
\begin{equation}
\ket{\theta,\phi} = \exp \left [ {{-i\phi\hat{S_z}}} \right ] \exp \left [ {{-i\theta\hat{S_y}}} \right ] \ket{S,S} 
\end{equation}
 Here, the phase space is parameterized by the
angles $\theta$ and $\phi$. To visualize phase space in terms of $\theta$ and $\phi$, we make use of  Husimi-Q function given by  
\begin{equation}
Q(\theta\,,\phi) = \frac{2S+1}{4\pi}\braket{\theta,\phi|\hat{\rho}_{{\rm atom}}|\theta,\phi}
\end{equation}
where $\hat{\rho}_{{\rm atom}}$ stands for the reduced atomic density operator. 
Roulet {\it et al.} \cite{Roulet2018ent} defined the shifted phase distribution by 
\begin{equation}
S(\phi|\hat{\rho}) = \int_{0}^{\pi}d\theta \sin(\theta)Q(\theta\,,\phi)-\frac{1}{2\pi}
\end{equation}\label{eq:Sphi_atom}
$S(\phi|\hat{\rho})$  is  zero for uniform (limit cycle) distribution and finite if there
is any phase reference present. Thus $S(\phi|\hat{\rho})$ may be considered as a measure of quantum synchronization. 

The Hammer Projection of the
Husimi-$Q$ function corresponding to the limit cycle  state is uniformly distributed
in the equatorial plane. Accordingly, $S(\phi|\hat{\rho})$ function 
becomes zero in this case.  The spin coherent state can also be expressed as \cite{LeeLoh2015}  

\begin{equation}
\ket{\theta\phi} = \left (\cos\frac{\theta}{2} \right )^{2} \left [\ket{-1} + \sqrt{2}\tan(\frac{\theta}{2})e^{i\phi}\ket{0} + {\tan}^2 \left (\frac{\theta}{2} \right ){e^{2i\phi}\ket{1}} \right]
\end{equation}

\subsection{Wehrl mutual information of the reduced two-mode field}
\label{sec:wehrl_mi}

To quantify correlations between the
two cavity modes, we compute the Wehrl mutual information
of the reduced two-mode field density matrix.  Starting from
the steady-state density matrix $\rho_{\mathrm{ss}}$, the reduced
field state is obtained by tracing out the atomic degrees of
freedom,
\begin{equation}
\rho_f
=
\mathrm{Tr}_{\mathrm{atom}}
\left(
\rho_{\mathrm{ss}}
\right).
\end{equation}

The corresponding single-mode reduced density matrices are
\begin{equation}
\rho_{f_1}
=
\mathrm{Tr}_2 \rho_f,
\qquad
\rho_{f_2}
=
\mathrm{Tr}_1 \rho_f.
\label{den}
\end{equation}

Following Ref.~\cite{PhysRevA.103.062222},
the two-mode Husimi distribution associated with $\rho_f$
is defined by
\begin{equation}
Q(\alpha_1,\alpha_2)
=
\langle
\alpha_1,\alpha_2
|
\rho_f
|
\alpha_1,\alpha_2
\rangle ,
\label{eq:QAB}
\end{equation}

where
\begin{equation}
|\alpha_1,\alpha_2\rangle
=
|\alpha_1\rangle
\otimes
|\alpha_2\rangle
\end{equation}

is the two-mode coherent state satisfying
\begin{equation}
a_j|\alpha_j\rangle
=
\alpha_j|\alpha_j\rangle,
\qquad
j\in\{1,2\}.
\end{equation}

Using the quadrature parametrization
\begin{equation}
\alpha_j
=
\frac{x_j+i p_j}{\sqrt{2}},
\label{alpha}
\end{equation}

the Husimi distribution satisfies the normalization condition
\begin{equation}
\int
\frac{
d^2\alpha_1\,d^2\alpha_2
}{
\pi^2
}
Q(\alpha_1,\alpha_2)
=
1,
\end{equation}

with
\begin{equation}
\frac{d^2\alpha_j}{\pi}
=
\frac{dx_j\,dp_j}{2\pi}.
\end{equation}

The single-mode Husimi marginals are obtained from
Eq.~\eqref{eq:QAB} through partial phase-space integration,
\begin{equation}
Q(\alpha_1)
=
\int
\frac{d^2\alpha_2}{\pi}
Q(\alpha_1,\alpha_2),
\end{equation}

\begin{equation}
Q(\alpha_2)
=
\int
\frac{d^2\alpha_1}{\pi}
Q(\alpha_1,\alpha_2).
\end{equation}

To characterize the phase-space structure of the individual cavity modes,
we consider the Wigner quasiprobability distributions associated with the
single-mode reduced density matrices introduced in Eq.~\eqref{den}.
For the $j$th cavity mode $(j=1,2)$, the Wigner function is defined as
\begin{equation}
W_j(x_j,p_j)
=
\frac{1}{\pi}
\int_{-\infty}^{\infty} d\xi_j\,
e^{2ip_j\xi_j}
\left\langle x_j-\xi_j
\middle|
\rho_{fj}
\middle|
x_j+\xi_j
\right\rangle .
\end{equation}
where the dimensionless quadratures are
$x_j=(a_j+a_j^\dagger)/\sqrt{2}$ and
$p_j=(a_j-a_j^\dagger)/(i\sqrt{2})$, related to the complex phase-space amplitude $\alpha_j$ through
Eq.~\eqref{alpha}, satisfying
$[x_j,p_j]=i$. The functions $W_1(x_1,p_1)$ and $W_2(x_2,p_2)$ therefore
represent the single-mode Wigner distributions of cavity modes $1$
and $2$, respectively, with the normalization
\begin{equation}
\int_{-\infty}^{\infty} dx_j
\int_{-\infty}^{\infty} dp_j\,
W_j(x_j,p_j)=1.
\end{equation}
Having characterized the phase-space structure of the individual cavity modes, we next quantify the correlations between the two modes using the Wehrl mutual information constructed from the Husimi-$Q$ distributions. The Wehrl entropy of the two-mode field state is then
defined as
\begin{equation}
S_W(\rho_f)
=
-
\int
\frac{
d^2\alpha_1\,d^2\alpha_2
}{
\pi^2
}
Q
(\alpha_1,\alpha_2)
\ln
Q
(\alpha_1,\alpha_2),
\label{eq:SWAB}
\end{equation}

while the corresponding single-mode Wehrl entropies are
\begin{equation}
S_W(\rho_{f_1})
=
-
\int
\frac{d^2\alpha_1}{\pi}
Q^{\rho_{f_1}}(\alpha_1)
\ln
Q^{\rho_{f_1}}(\alpha_1),
\label{eq:SWA}
\end{equation}

\begin{equation}
S_W(\rho_{f_2})
=
-
\int
\frac{d^2\alpha_2}{\pi}
Q^{\rho_{f_2}}(\alpha_2)
\ln
Q^{\rho_{f_2}}(\alpha_2).
\label{eq:SWB}
\end{equation}

The Wehrl mutual information is defined by
\begin{equation}
I_W(f_1:f_2)
=
S_W(\rho_{f_1})
+
S_W(\rho_{f_2})
-
S_W(\rho_f).
\label{eq:IW}
\end{equation}
Equation~\eqref{eq:IW} is the Wehrl analogue of the quantum
mutual information. It is a measure of the deviation of the joint
Husimi distribution from its product separable form of 
single-mode Husimi marginals.  In terms of the Husimi
distributions, the Wehrl relative entropy between the full
distribution and the factorized distribution is given by
\begin{align}
I_W(f_1:f_2)
&=
\int
\frac{
d^2\alpha_1\,d^2\alpha_2
}{
\pi^2
}
Q(\alpha_1,\alpha_2)
\nonumber\\
&\times
\ln
\left[
\frac{
Q(\alpha_1,\alpha_2)
}{
Q^{\rho_{f_1}}(\alpha_1)
Q^{\rho_{f_2}}(\alpha_2)
}
\right].
\label{eq:IW_relative}
\end{align}
Since the Husimi distribution is non-negative and normalized,
the Wehrl mutual information satisfies
\begin{equation}
I_W(f_1:f_2)\ge0,
\end{equation}

with equality only when the joint Husimi distribution
factorizes,
\begin{equation}
Q(\alpha_1,\alpha_2)
=
Q^{\rho_{f_1}}(\alpha_1)
Q^{\rho_{f_2}}(\alpha_2).
\end{equation}

For the numerical evaluation, the Husimi distributions are
constructed directly from the Fock-basis matrix elements of
the reduced density matrices.  For the single-mode reduced
states,
\begin{equation}
Q(\alpha)
=
e^{-|\alpha|^2}
\sum_{n,m}
\frac{
(\rho_j)_{nm}
(\alpha^*)^n
\alpha^m
}{
\sqrt{n!\,m!}
},
\qquad
j\in\{f_1,f_2\},
\label{eq:Qsingle}
\end{equation}

while the two-mode Husimi distribution is
\begin{align}
Q(\alpha_1,\alpha_2)
&=
e^{-|\alpha_1|^2-|\alpha_2|^2}
\nonumber\\
&\times
\sum_{n_1,n_2,m_1,m_2}
\frac{
(\rho_f)_{(n_1n_2)(m_1m_2)}
(\alpha_1^*)^{n_1}
\alpha_1^{m_1}
(\alpha_2^*)^{n_2}
\alpha_2^{m_2}
}{
\sqrt{
n_1!\,
n_2!\,
m_1!\,
m_2!
}
}.
\label{eq:Qjoint}
\end{align}

The phase-space integrations in
Eqs.~\eqref{eq:SWAB}--\eqref{eq:SWB}
are evaluated numerically on finite quadrature grids using
the measure
$d^2\alpha/\pi = dx\,dp/(2\pi)$.
For each steady-state density matrix,
the quantities
$S_W(\rho_{f_1})$,
$S_W(\rho_{f_2})$,
$S_W(\rho_f)$,
and
$I_W(f_1:f_2)$
are computed directly from the corresponding Husimi
distributions.
To compare heterodyne-accessible intermode correlations
with the phase-space synchronization structure,
the Wehrl mutual information
$I_W(f_1:f_2)$
is analyzed together with the synchronization measure
defined in Eq.~\eqref{eq:Sphi_def}.
While the synchronization measure probes the angular
modulation structure of the generalized phase-space
distribution,
$I_W(f_1:f_2)$
quantifies the deviation of the joint Husimi distribution from
a factorized product of single-mode Husimi marginals.
Both quantities are evaluated throughout the parameter scans
over
$\kappa_2/\kappa_1$.

\section{Results and Discussion}\label{Discussion}

We  use joint atom-field basis states in which field basis are the product  Fock states of photon numbers for the two modes. We simulate the Lindblad dynamics using the quantum-trajectory method and take the trajectory-averaged density matrix
$\rho(t_{\rm f})$ at sufficiently long time.
Each cavity mode is truncated to $N$ Fock states, with convergence verified
by requiring $P_{1,2}(N-1)<10^{-7}$ and checking stability upon increasing $N$. We then obtain reduced atomic and two-mode field steady-state density matrices by tracing over the field and atomic states, respectively, of the joint density matrix. We present numerical results that establish the occurrence of limit cycles and quantum synchronization in field and atomic sub-systems depending on the system parameters. 

Before we present our results, it is worthwhile to note the basic difference between classical and quantum limit cycles. 
In classical nonlinear dynamics, a limit cycle is characterized by the stabilization of the oscillation amplitude while the phase remains free. In an open quantum system as considered here, this notion must be reformulated at the level of the steady-state density matrix. A quantum limit cycle is identified by the coexistence of
\begin{equation}
\langle \hat{a}_i^\dagger \hat{a}_i \rangle_{\mathrm{ss}} > 0,
\qquad
\langle \hat{a}_i \rangle_{\mathrm{ss}} = 0,
\label{eq:limitcycle_def}
\end{equation}
for each mode $i=1,2$. The finite steady-state occupations $\langle \hat{a}_i^\dagger \hat{a}_i \rangle_{\mathrm{ss}}$ indicate that the field amplitude is stabilized, while the vanishing first moments $\langle \hat{a}_i\rangle_{\mathrm{ss}}$ show that no definite phase is selected. The vanishing of $\langle \hat{a}_i\rangle_{\mathrm{ss}}$ follows from the continuous $U(1)$ phase symmetry of the Liouvillian. In the absence of an external phase reference, the steady state is invariant under
\begin{equation}
\rho_{\mathrm{ss}}
\rightarrow
e^{i\phi(\hat{a}_1^\dagger\hat{a}_1+\hat{a}_2^\dagger\hat{a}_2)}
\rho_{\mathrm{ss}}
e^{-i\phi(\hat{a}_1^\dagger\hat{a}_1+\hat{a}_2^\dagger\hat{a}_2)} .
\end{equation}
This implies
\begin{equation}
\langle \hat{a}_i\rangle_{\mathrm{ss}}
=
e^{i\phi}\langle \hat{a}_i\rangle_{\mathrm{ss}}
\end{equation}
for arbitrary $\phi$ and therefore enforces $\langle \hat{a}_i\rangle_{\mathrm{ss}}=0$.

  \begin{figure}[H]
    \centering
    
    \includegraphics[
        width=0.8\textwidth,
        keepaspectratio
    ]{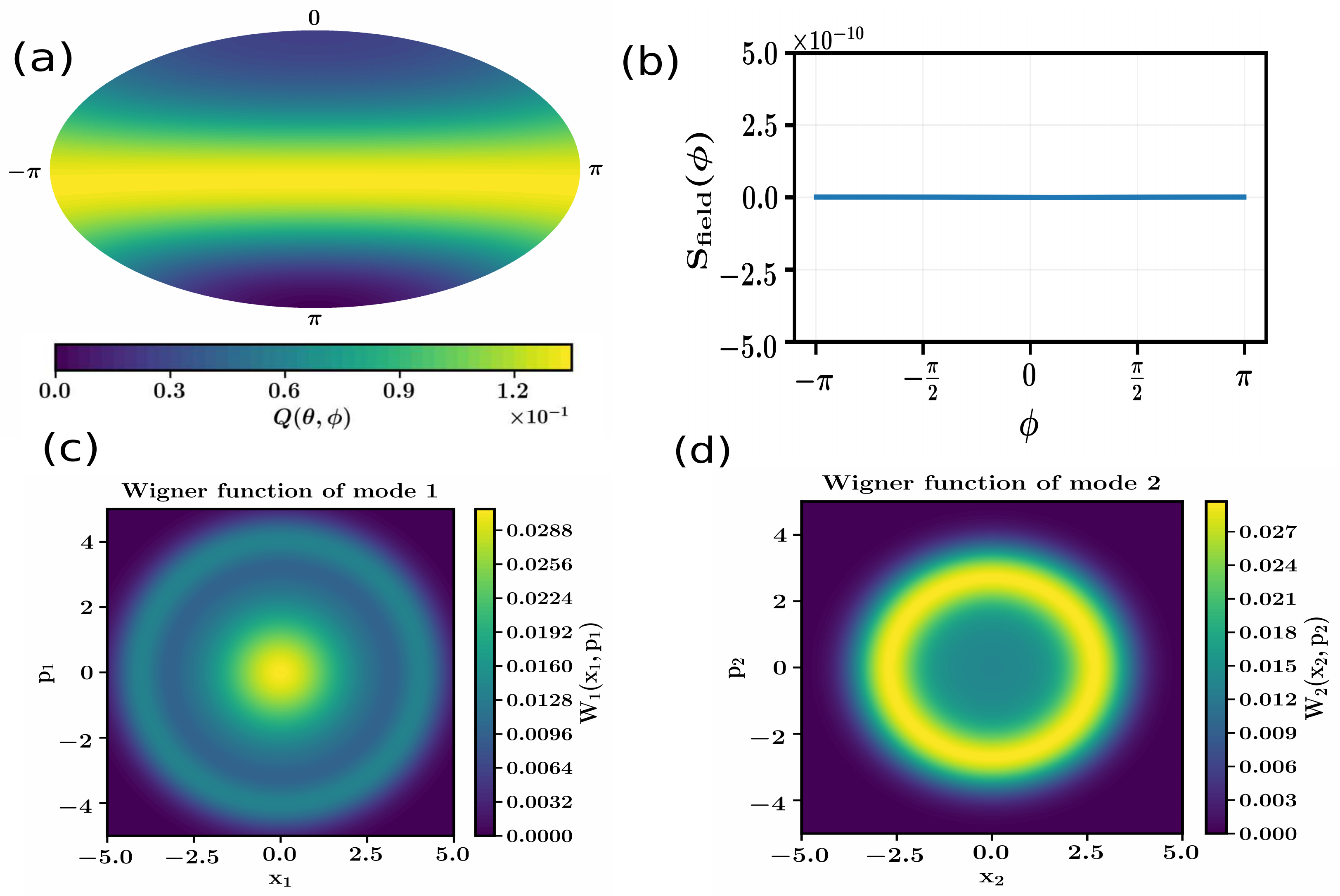}
  
  \caption{
    (a) Plotted is $Q_{\rm field}(\theta,\phi)$ for no cavity drives i.e.
    $\Omega_1 = \Omega_2 = 0$, $\Omega = 0.45\Gamma_d$
    $\kappa_1=\kappa_2 = 10^{-4}\Gamma_d$, 
    $g_1=g_2=0.5\Gamma_d$, 
    $\Delta_d = 100\Gamma_d$, $\Delta_a = 0.01\Gamma_d$, $\Delta_c = -0.01\Gamma_d$, $\delta_1 = -0.001$,  $\delta_2 = 0.001$ and 
    $\gamma_1=\gamma_2 = 1\times10^{-2}\Gamma_d$ and the subplot  
    (b) shows the corresponding $S_{field}$. Subplots (a) and (b) reveal a cavity-field limit cycle.
    Subplots (c) and (d) show the Wigner quasiprobability distributions 
    $W_1(x_1,p_1)$ and 
    $W_2(x_2,p_2)$ of the cavity field mode 1 and mode 2, respectively. 
     The finite-radius annular Wigner distributions centered away from the vacuum point confirm the existence of nonzero cavity-field amplitudes with phase diffusion as a characteristic feature  of a limit cycle. 
    The conditions 
    $\langle \hat{a}_1 \rangle=\langle \hat{a}_2 \rangle = 0$ together with 
    $\langle \hat{a}_1^\dagger\hat{a}_1\rangle =2.712$ and 
    $\langle \hat{a}_2^\dagger\hat{a}_2\rangle = 1.976$ further establish the occurrence of the  limit cycle.
    }

    \label{fig:husimi_wigner_compare}
\end{figure}

\begin{figure}[H]
    \centering
    
    \includegraphics[
        width=0.8\textwidth,
        keepaspectratio
    ]{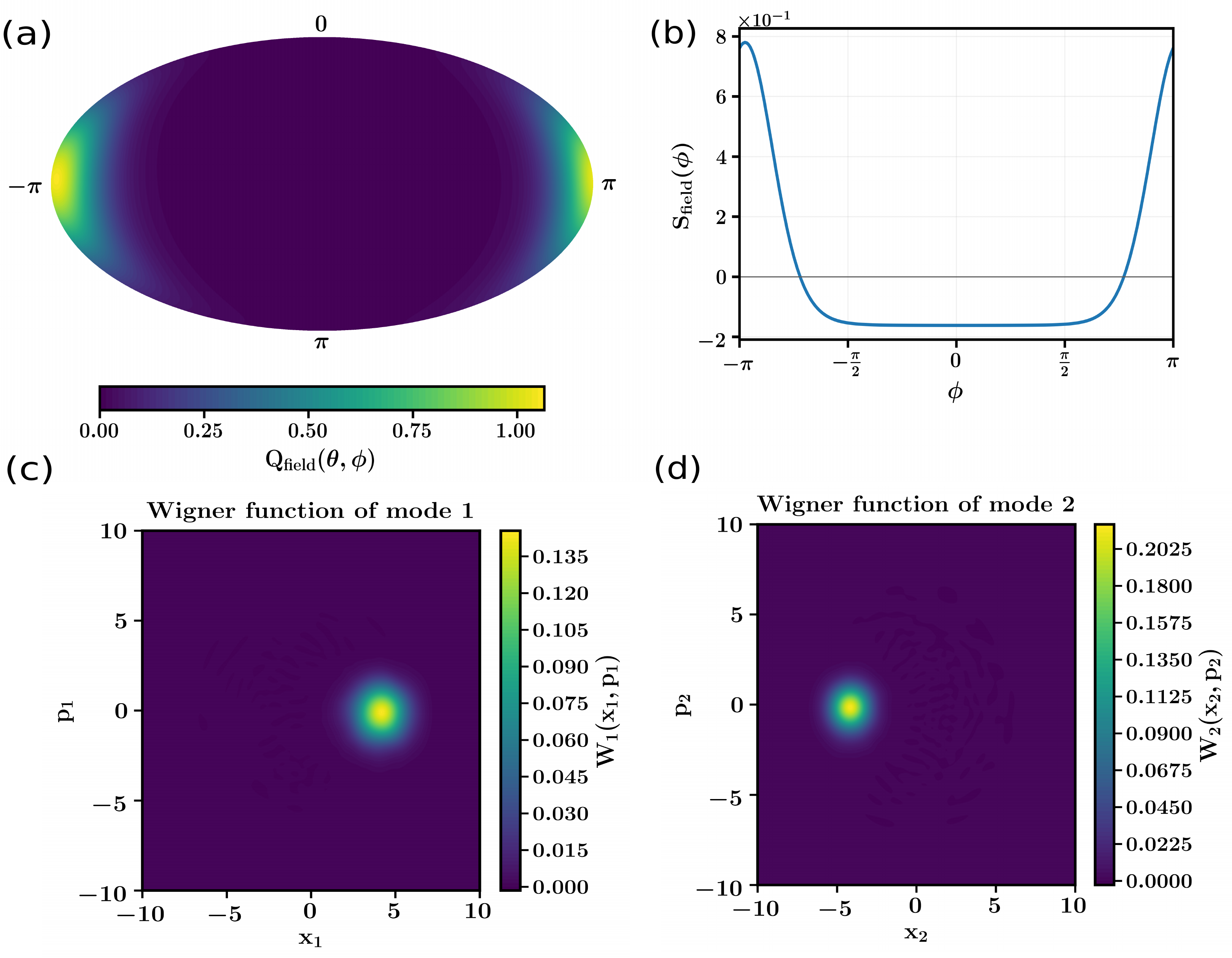}
\caption{
$Q_{\rm field}$ (a) and Wigner distributions (c,d) for 
$\Omega_1=\Omega_2=0.005\Gamma_d$, with all other parameters remaining fixed as in Fig.~2. 
Subplot (b) shows the corresponding 
$S_{\rm field}(\phi)$. 
Subplots (c) and (d) display 
$W_1(x_1,p_1)$ and $W_2(x_2,p_2)$, respectively. 
The  phase localization in (a) along with the crescent-shaped $W_1(x_1,p_1)$ and $W_2(x_2,p_2)$ in  (c) and (d), indicates that the  cavity drives force the system to select a definite phase by distorting the phase-diffused limit cycle, leading to the emergence of an out-of-phase two-mode synchronized field state. 
Here 
$\langle \hat{a}_1^\dagger\hat{a}_1\rangle=9.07$, 
$\langle \hat{a}_2^\dagger\hat{a}_2\rangle=8.97$, 
$\langle \hat{a}_1\rangle= 2.90 -0.10i$ and 
$\langle \hat{a}_2\rangle=-2.93 -0.14i$.
}
\label{fig:field_outphase_drive}
\end{figure}

\begin{figure}[H]
    \centering
    
    \includegraphics[
        width=0.8\textwidth,
        keepaspectratio
    ]{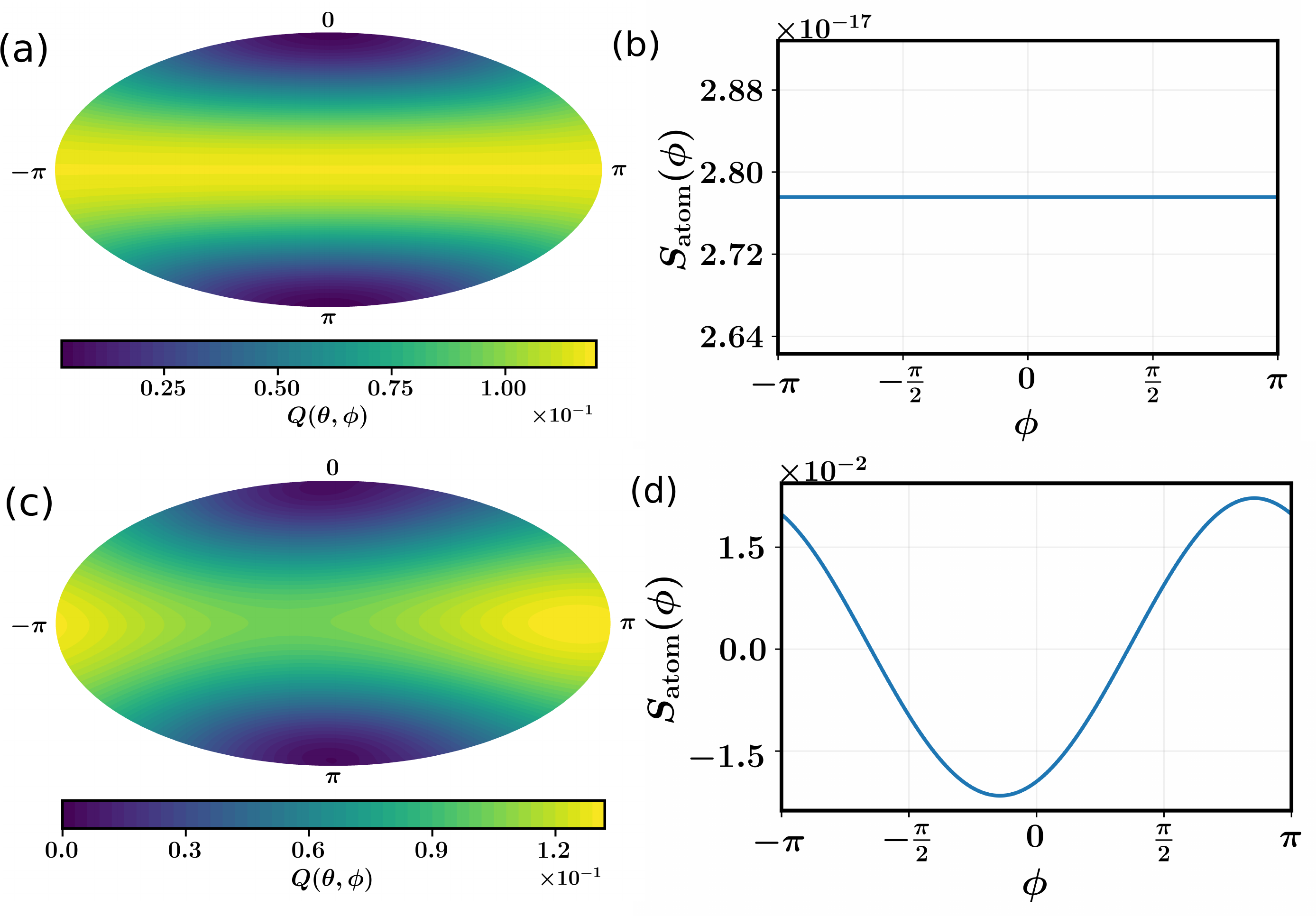}

\caption{
Atomic Husimi-$Q$ distribution (a,c)  and the corresponding  $S_{atom}$ (b,d) for $\Omega = 0.45\Gamma_d$, $\kappa_1=\kappa_2 = 10^{-4}\Gamma_d$, $g_1=g_2=0.5\Gamma_d$, $\Delta_d = 100\Gamma_d$ and $\gamma_1=\gamma_2 = 1\times10^{-2}\Gamma_d$. Subplots (a) and (b) correspond to the no cavity-drive case, $\Omega_1=\Omega_2=0$, where $Q(\theta,\phi)$ remains nearly phase-independent and the corresponding $S_{\mathrm{atom}}(\phi)$ is essentially flat, indicating a limit-cycle. Subplots (c) and (d) correspond to finite cavity drives, $\Omega_1=\Omega_2=0.005\Gamma_d$, where the atomic phase-space distribution becomes  localized and the corresponding $S_{atom}$ develops pronounced structure with enhanced weight near $\phi=\pm\pi$, demonstrating the emergence of atomic out-of-phase synchronization. The results displayed in this and the previous figure show that by increasing the cavity  drive strengths to a small level, it is possible to achieve synchronization in both the cavity field and the atomic subsystems.
}
\label{fig:atomic_husimi_drive}
\end{figure}

\begin{figure}[H]
    \centering
    \includegraphics[width=\linewidth]{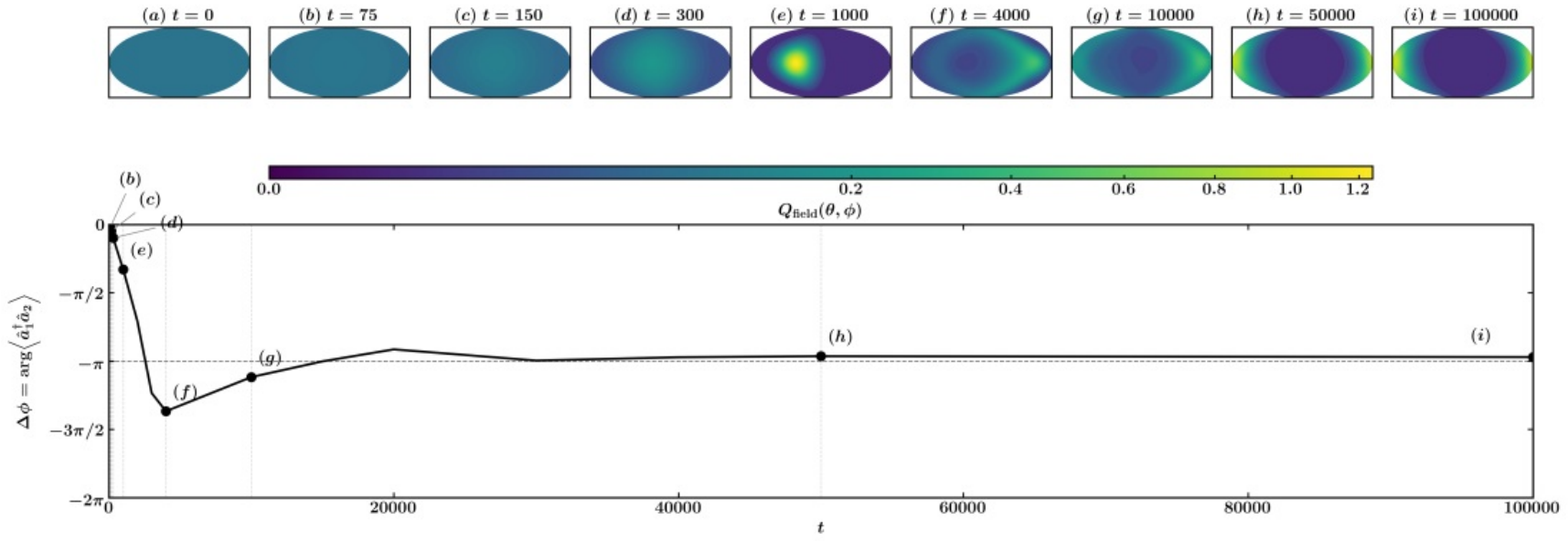}
    \caption{
    Dynamical development of relative-phase locking in the two-mode field.
    The upper panels show the two-mode field Husimi-$Q$ distribution
    $Q_{\mathrm{field}}(\theta,\phi;t)$ at representative evolution times. The lower panel shows the relative
    phase
    $\Delta\phi(t)=
    \arg\langle\hat a_1^{\dagger}\hat a_2\rangle_t$,
    with the marked points corresponding to the Husimi snapshots above.
    }
    \label{fig:husimi_phase_dynamics}
\end{figure}

\begin{figure}[H]
    \centering
    
    \includegraphics[
        width=0.8\textwidth,
        keepaspectratio
    ]{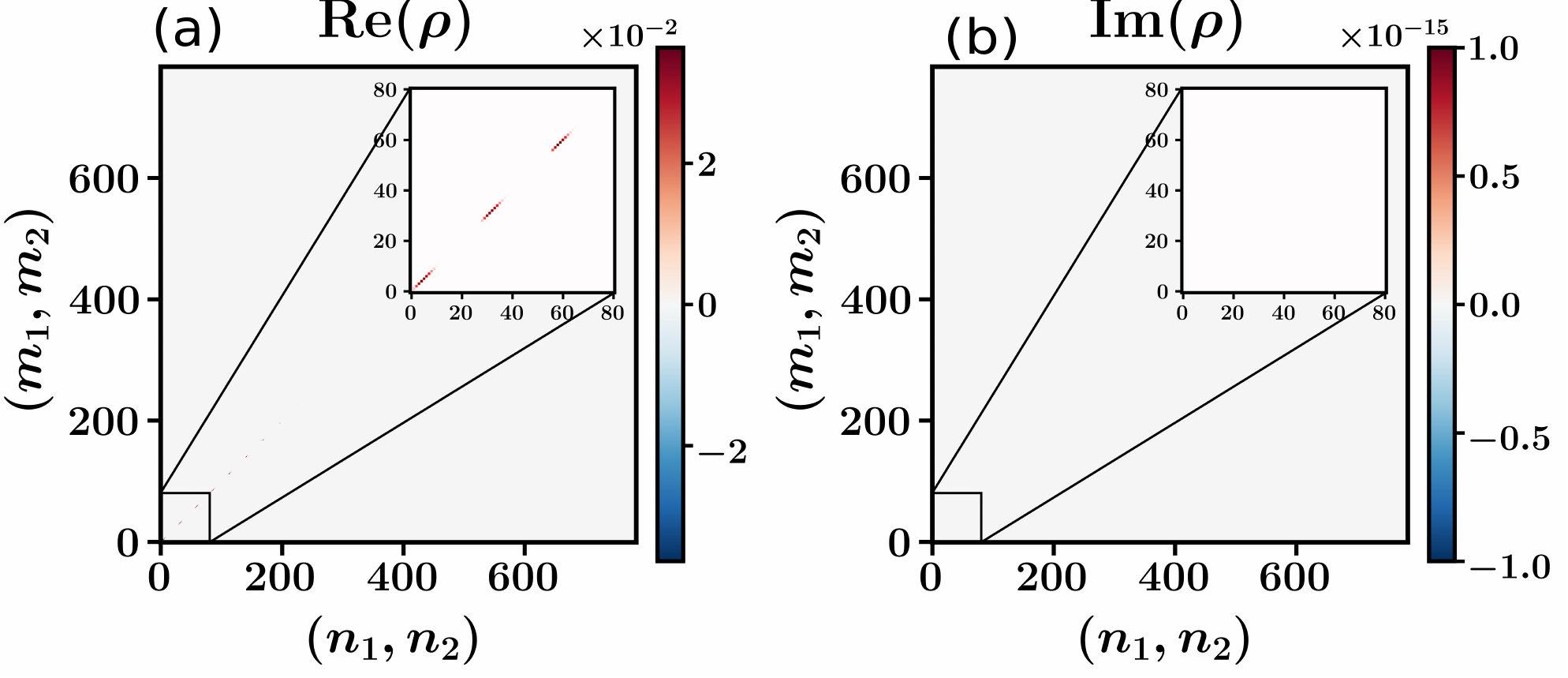}

\caption{
Real (a) and imaginary (b) parts of the reduced two-mode cavity-field density matrix elements $[\rho_{\rm f}]_{i j}$, where the indices $i \equiv (n_1, n_2)$ and $j \equiv (m_1, m_2) $ stand for the joint two-mode Fock basis $\ket{n_1, n_2}$ and $\ket{m_1, m_2}$, respectively; for the parameters remaining the same as in Figs.~\ref{fig:husimi_wigner_compare}(a)
and \ref{fig:husimi_wigner_compare}(b). The subscripts `1' and `2' refer to the mode 1 and 2, respectively.   The photon-number pairs $(n_1, n_2)$ and $(m_1, m_2)$ are mapped to flattened basis indices $i$ and $j$ with
$i=n_1N+n_2$ and $j=m_1N+m_2$. 
 The density matrix is
dominated by diagonal terms (photon-number populations), while the
off-diagonal terms and the imaginary components are negligible,
indicating an incoherent limit-cycle state without phase locking between
the two cavity modes.
}
\label{fig:field_density_limitcycle}
\end{figure}

\begin{figure}[H]
    \centering
    
    \includegraphics[
        width=0.8\textwidth,
        keepaspectratio
    ]{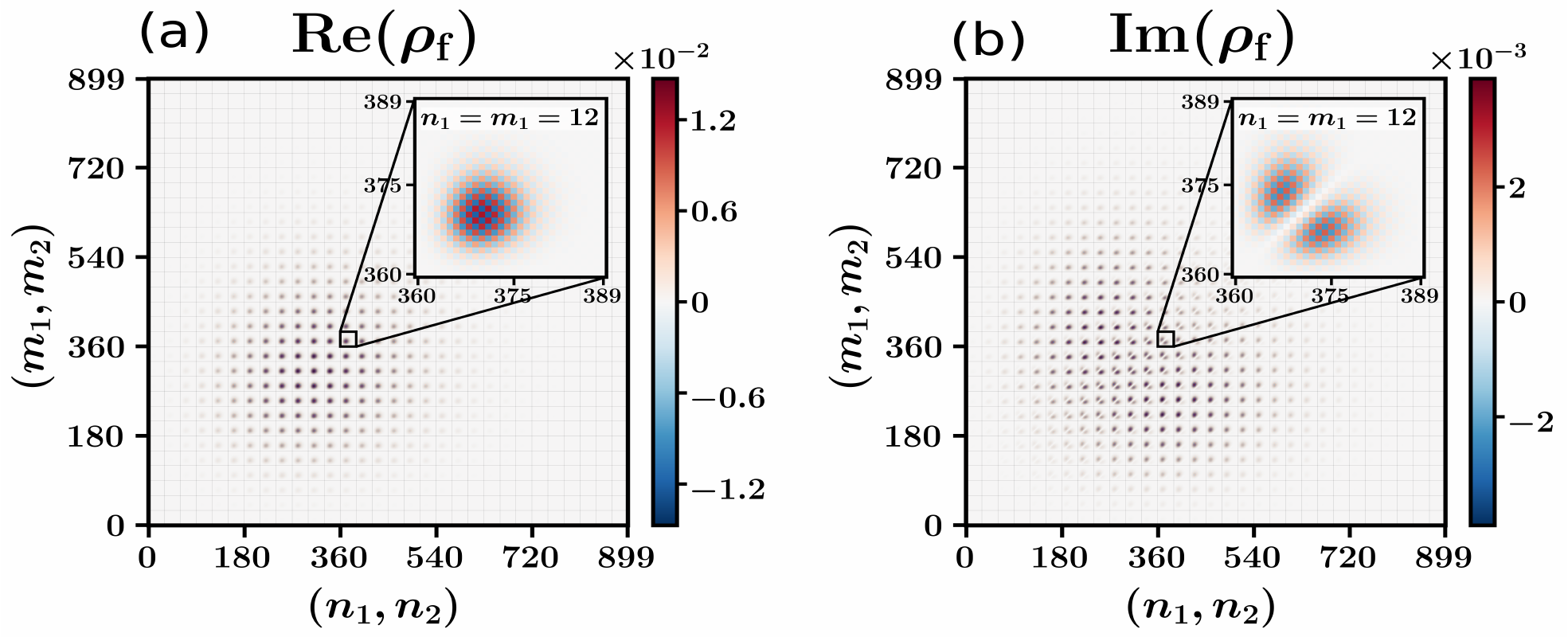}

\caption{
Same as in the previous figure but for the parameters of  the out-of-phase synchronized regime  of Fig.3. 
Compared with the limit-cycle state in Fig.~\ref{fig:field_density_limitcycle}, finite off-diagonal elements appear in the joint Fock basis, showing the buildup of phase-sensitive field coherences. 
}
\label{fig:field_density_outphase}
\end{figure}

\begin{figure}[H]
    \centering
    
    \includegraphics[
        width=0.8\textwidth,
        keepaspectratio
    ]{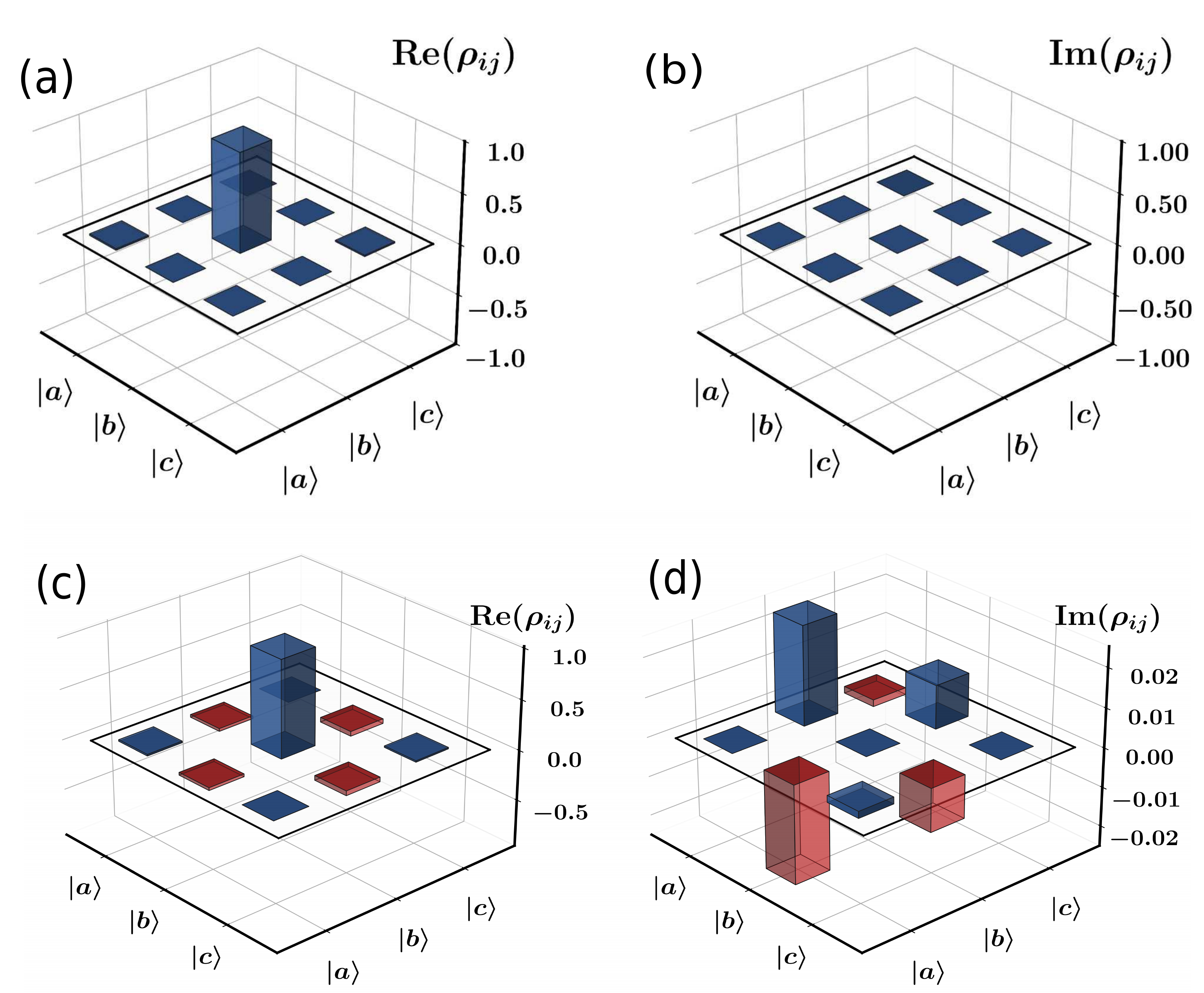}

\caption{
Real (a) and imaginary (b) parts of the reduced atomic density-matrix elements  in the $\{|a\rangle,|b\rangle,|c\rangle\}$ basis. 
The population $\rho_{bb}$ in state $\ket{b}$ is dominant and all off-diagonal terms vanish. 
Subplots (c) and (d) show the corresponding density matrix elements in the synchronized regime under finite cavity driving, where nonzero off-diagonal terms (coherences) emerge. 
The appearance of finite  coherences in Subplot (d), together with the redistribution of the real-part matrix elements in Subplot (c), is due to the transition from a limit-cycle  to a phase-locked synchronized atomic state.
}
\label{X}
\end{figure}

\begin{figure}[H]
    \centering
    \begin{minipage}[t]{\textwidth}
        \centering
        \includegraphics[width=\linewidth]{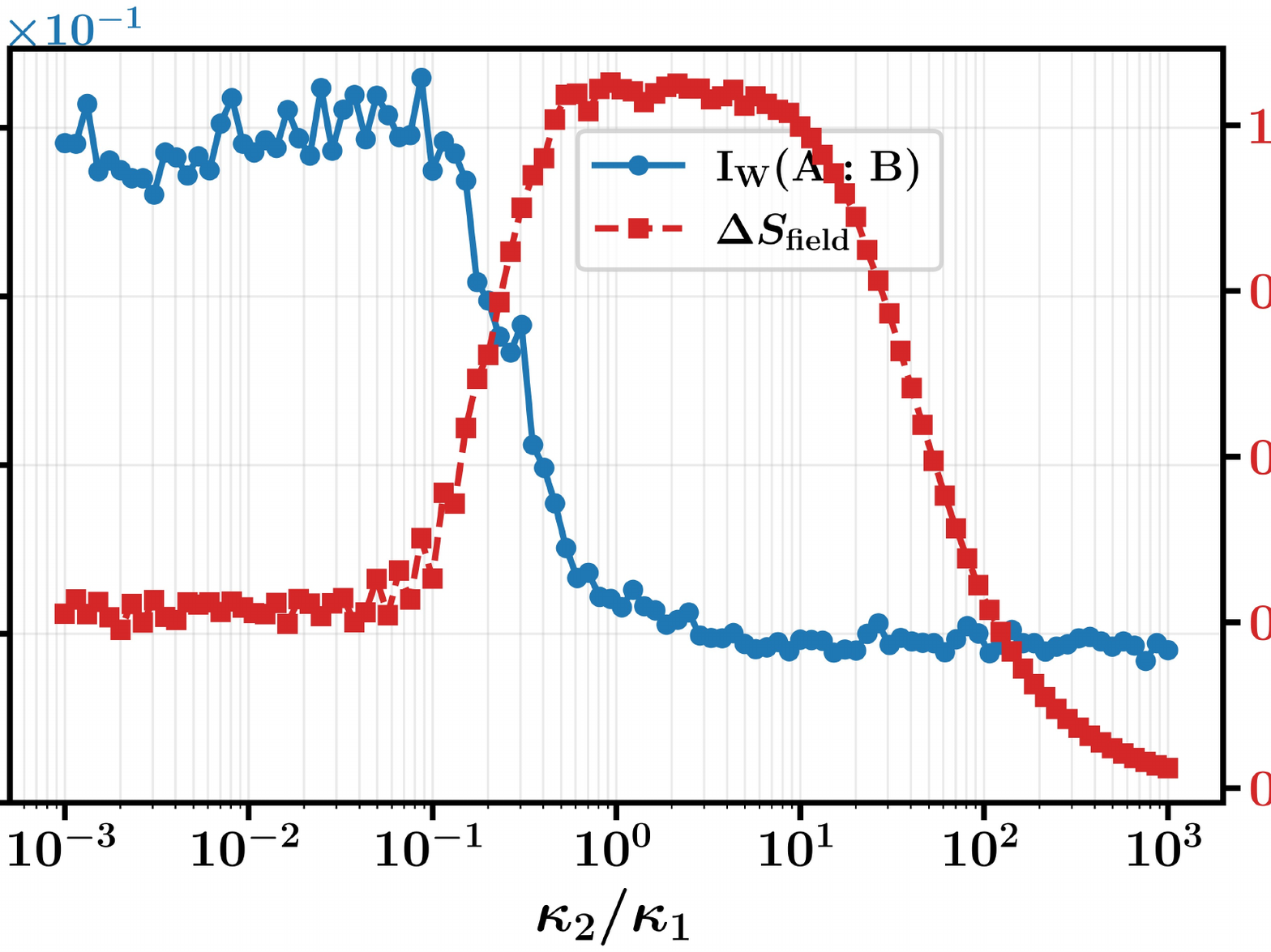}
    \end{minipage}
 \caption{%
        \textbf{(a)} Wehrl mutual information $I_W(A{:}B)$ (blue, left axis)
        and $\Delta S_{\mathrm{field}} = \max [S_{\mathrm{field}}]
        - \min [S_{\mathrm{field}}]$ (red, right axis) versus cavity-loss asymmetry
        $\kappa_2/\kappa_1 \in [10^{-3},10^{3}]$ with atomic decay rates
        held fixed at ($\gamma_1 = \gamma_2 = 10^{-2}\Gamma_d$).
        Increasing cavity loss imbalance again reduces $I_W$ while
        enhancing phase-selective localization measured by
        $\Delta S_{\mathrm{field}}$.
    }
    \label{fig:wehrl_sync}
\end{figure}

\subsection{Quantum limit cycle and phase locking in the bimodal cavity field}
 Now, we discuss our numerical results presented in the Fig.2-9. In Figs. 2 and 3 we show the results pertaining to the two-mode field limit cycle and quantum synchronization. Figure 4 displays the results for atomic spin. The dynamical field phase is shown in Fig.5. For the chosen input parameters of these figures, we obtain convergent results if we truncate  the total photon number to 60 or above with each mode having at least 30 photons. The reduced field density matrix elements are depicted in Figs. 6 and 7; while the reduced atomic density matrix elements are displayed in Fig. 8. The results of Wehrl mutual information are presented in Fig.9. In what  follows, we discuss these figures one by one. 

Figure~\ref{fig:husimi_wigner_compare} shows the two-mode field Husimi-$Q$ distribution $Q_{\rm field}(\theta, \phi)$ (a) and the corresponding phase marginal $S_{\mathrm{field}}(\phi)$ (b) in the case of no cavity drive, that is, $\Omega_1=\Omega_2=0$. $Q_{\rm field}(\theta, \phi)$ in Fig.~\ref{fig:husimi_wigner_compare}(a) is nearly independent of $\phi$ and the corresponding $S_{\mathrm{field}}(\phi)$ in Fig.~\ref{fig:husimi_wigner_compare} (b) is almost flat, indicating  that the field is not phase-locked. The Wigner functions in Figs.~\ref{fig:husimi_wigner_compare}(c) and \ref{fig:husimi_wigner_compare}(d) show finite-radius annular structures, suggesting that the two field modes are populated and amplitude-stabilized, but their phases remain diffusive. This interpretation is supported by the vanishing first moments, $\langle \hat{a}_1\rangle=\langle \hat{a}_2\rangle=0$, together with finite photon occupations, $\langle \hat{a}_1^\dagger\hat{a}_1\rangle=2.712$ and $\langle \hat{a}_2^\dagger\hat{a}_2\rangle=1.976$.
Thus Fig.~\ref{fig:husimi_wigner_compare} provides direct evidence of a two-mode field limit cycle. Physically, the limit-cycle field state is a phase-averaged two-mode state with finite radius in phase space. This gives a ring-like quasi-probability distribution: The amplitude is stabilized, but the phase diffuses freely. Note that, the steady-state in this case is reached primarily due to the competition between the atomic drive and atomic damping, since there is no cavity drive and cavity damping rate is much smaller than the atomic decay rate.  

Figure~\ref{fig:field_outphase_drive} shows the effect of small cavity drive, $\Omega_1=\Omega_2=0.005\Gamma_d$, with all other parameters remaining fixed as in Fig.~\ref{fig:husimi_wigner_compare}. $Q_{\mathrm field}(\theta, \phi)$  in Fig.~\ref{fig:field_outphase_drive}(a) is no longer uniform in $\phi$ coordinate. $S_{\mathrm{field}}(\phi)$ in Fig.~\ref{fig:field_outphase_drive}(b) has enhanced weight near $\phi=\pm\pi$ and a minimum near $\phi=0$,  evidencing an out-of-phase synchronization. The first moments, $\langle \hat{a}_1\rangle= 2.90 -0.10i$ and $\langle \hat{a}_2\rangle=-2.93 -0.14i$, show that the two field amplitudes lie mainly on opposite sides of the real axis, with $\langle \hat{a}_1\rangle$ having a positive real part and $\langle \hat{a}_2\rangle$ having a negative real part. Therefore, the relative phase between the two modes is close to $\pi$, consistent with out-of-phase locking. The Wigner functions in Figs.~\ref{fig:field_outphase_drive}(c) and \ref{fig:field_outphase_drive}(d) remain close to annular structures but become weakly distorted due to the cavity drives. Therefore, a  weak drive does not completely destroy the limit-cycle background, but  it can force the system to select a relative phase to produce an out-of-phase synchronized branch. Since, in this case, cavity drive is much smaller than the atomic drive, the cavity photons are mostly generated due to the atomic drive, nevertheless extremely weak classical fields driving the two cavity modes in the presence of relatively stronger atomic  drive can lead to U(1) symmetry breaking and thereby leading to the two-mode field quantum synchronization. 

Figure~\ref{fig:atomic_husimi_drive} shows the results for  the atomic subsystem. Figs.~\ref{fig:atomic_husimi_drive}(a) and \ref{fig:atomic_husimi_drive}(b) correspond to the no cavity-drive case,  $\Omega_1=\Omega_2=0$.  The atomic Husimi distribution $Q_{\mathrm{atom}}(\theta,\phi)$ is nearly phase-independent and the corresponding $S_{\mathrm{atom}}(\phi)$ is flat, showing  that the atomic state is also in a limit-cycle-like regime without any phase locking.  Figs.~\ref{fig:atomic_husimi_drive}(c) and \ref{fig:atomic_husimi_drive}(d) show that  finite cavity drives can lead to atomic synchronization modulations in $S_{\mathrm{atom}}(\phi)$. 

Figure~\ref{fig:husimi_phase_dynamics} shows the dynamical evolution of the
two-mode field Husimi-$Q$ distribution together with the relative phase
extracted from the first-order intermode coherence,
\begin{equation}
    \Delta\phi(t)
    =
    \arg\!\left[
        \left\langle
        \hat a_1^{\dagger}\hat a_2
        \right\rangle_t
    \right].
    \label{eq:relative_phase}
\end{equation}
At the initial time, the field distribution is phase delocalized and the
relative phase is not well defined because the intermode coherence is
vanishingly small. As the system evolves, a phase-dependent
structure develops in the Husimi distribution. During the transient regime,
the relative phase passes through the intermediate values. Subsequently, the phase approaches $\Delta\phi\simeq-\pi$ and remains close
to this value at long times. The simultaneous development of a structured
two-mode Husimi distribution and stabilization of the intermode
phase near $-\pi$ is consistent with the dynamical formation of an
anti-phase-locked configuration. Importantly, the transient evolution shows
that the approach to the long-time phase-locked regime proceeds through
intermediate relative phases.

Figure~\ref{fig:field_density_limitcycle} shows the reduced two-mode cavity-field density matrix in the limit-cycle regime. The matrix elements are written in the joint Fock basis as
$\rho_{(n_1,n_2),(m_1,m_2)}
=\langle n_1,n_2|\rho_f|m_1,m_2\rangle$.  The density matrix is dominated by the diagonal terms which represent photon number populations, while the imaginary components  vanish. This structure is consistent with a phase-diffusive limit-cycle state: The two cavity modes have finite steady-state occupation, but without any relative phase. The density-matrix structure changes once phase locking is established. In the out-of-phase synchronized regime, shown in Fig.~\ref{fig:field_density_outphase}, finite off-diagonal elements appear. These matrix elements represent coherences between different two-mode photon-number states and provide the microscopic origin of the phase-dependent modulation of $S_{\mathrm{field}}(\phi)$. The corresponding phase marginal is localized near $\phi=\pm\pi$, showing that these coherences support preferential out-of-phase locking. Compared with the limit-cycle state, the off-diagonal structure is more pronounced, indicating that the steady state is no longer an incoherent mixture of joint Fock populations. The buildup of these phase-sensitive coherences is consistent with the shift of the phase marginal toward $\phi=\pm\pi$.  Figure~\ref{X} shows the corresponding reduced atomic density matrix. In the limit-cycle regime, the atomic state is dominated by the population term $\rho_{bb}$ and the coherences are negligible. This means that the atomic state does not carry a well-defined  phase. In the synchronized regime, finite off-diagonal atomic coherences appear. These coherences show that the atomic subsystem also develops phase sensitivity when the cavity field becomes synchronized. Thus the atomic density matrix vindicates the same conclusion as drawn from the analysis of the atomic Husimi distribution: Synchronization is accompanied by the appearance of atomic coherences.

Finally, Fig.~\ref{fig:wehrl_sync} compares the Wehrl mutual information $I_W(A{:}B)$ with the phase-locking contrast $\Delta S_{\mathrm{field}}=\max[S_{\mathrm{field}}]-\min[S_{\mathrm{field}}]$. In this figure, the scan is performed as a function of the cavity-loss asymmetry $\kappa_2/\kappa_1$, with equal atomic decay rates. As $\kappa_2/\kappa_1$ increases, $I_W(A{:}B)$ decreases, whereas $\Delta S_{\mathrm{field}}$ increases. It is interesting to note that as the Wehrl information decreases, the phase-locking contrast increases. These results show that the two-mode correlation measured by $I_W(A{:}B)$ is not identical to phase synchronization. In this system, stronger phase localization can occur even when the Wehrl mutual information decreases. Thus the data indicate a trade-off between Wehrl mutual information and phase-selective synchronization.

In summary,  our results reveal several interesting aspects of quantum synchronization in a driven dissipative, strongly interacting atom-cavity coupled system. Without cavity driving, the two-mode field and the atomic subsystem form phase-diffusive limit-cycle states. A weak cavity drive  breaks the phase symmetry and produces out-of-phase synchronization. The density matrices show that phase locking is accompanied by the appearance of off-diagonal terms or coherences. The Wehrl mutual information analysis further shows that phase synchronization and total two-mode phase-space correlation are distinct diagnostics: in the present parameter regimes, enhanced phase localization is accompanied by reduced Wehrl mutual information.

\section{Conclusions}\label{conclusion}
\label{Conclusions}
In conclusion, we have demonstrated quantum synchronization in a driven-dissipative atom-cavity system where an effective spin-$1$ atomic ensemble mediates the interaction between two cavity modes. Starting from the effective Hamiltonian and master equation, we have identified both phase-diffused limit-cycle states and phase-localized synchronized states in the fields as well as in the atoms within the same physical platform.

In the absence of cavity driving, the field and atomic subsystems form limit-cycle states: the steady-state occupations are finite, but no definite phase is selected. This is confirmed by the annular Wigner distributions, the nearly phase-independent generalized two-mode Husimi distribution, and the flat phase marginal $S_{\mathrm{field}}(\phi)$. When weak cavity drives are applied, the phase-diffusive field develops a preferred relative phase. For the parameters considered here, weak equal cavity drives produce an out-of-phase synchronized state, characterized by enhanced phase weight near $\phi=\pm\pi$. The development of synchronization is also accompanied by nonzero off-diagonal elements in the reduced density matrices, reflecting the development of phase coherence.

The density-matrix results support this phase-space picture. In the limit-cycle regime, the reduced field density matrix is dominated by diagonal photon-number populations. In the synchronized regimes, phase-sensitive off-diagonal coherences appear in the joint two-mode Fock basis and generate the angular modulation of $S_{\mathrm{field}}(\phi)$. The atomic density matrix shows a similar buildup of coherences, indicating that the atomic and field synchronization arise from the same atom-mediated Raman dynamics.

The photon-sector analysis further supports the use of the generalized field Husimi construction. In the parameter regimes studied here, the reduced field state is distributed over several total-photon-number sectors rather than being well represented by a single fixed-\(N\) shell. The synchronization phenomena therefore arise from the collective phase structure of the occupied Schwinger sectors, making a multi-sector phase-space description necessary.

Finally, the comparison with Wehrl mutual information shows that phase synchronization and total two-mode phase-space correlation are distinct diagnostics. Stronger phase localization can occur while the Wehrl mutual information decreases, showing that synchronization is not simply equivalent to increased mode-mode correlations. Overall, our results show how cavity-mediated coherent Raman coupling, cavity driving and dissipation can be used to transform a quantum limit cycle into phase synchronized states in an open atom-cavity system.

\vspace{0.5cm}

\noindent 

\section*{Acknowledgments}

KH gratefully acknowledges the Indian Association for the Cultivation of Science for fellowship support during her Master's thesis, during which this work was initiated. KH also thanks Soumalya Haldar for assistance in preparing the schematic diagram using Blender. SG and BD acknowledge support from the Department of Science and Technology, Government of India, under the National Quantum Mission in the area of Quantum Sensing and Metrology.

\subsection*{\Large{Appendix}}
\label{subsec: adia}

Adiabatic elimination is a technique \cite{Brion2007} used to simplify the dynamics of a multi-level system by effectively eliminating certain fast variables. It is commonly employed when there is a significant separation in timescales between different subsystems or degrees of freedom. Let's discuss the application of adiabatic elimination in a four-level system.\\
To apply adiabatic elimination, we assume that the dynamics of certain levels are much faster than the others. Let's assume that the level $\ket{4}$ is the fast levels, while $\ket{1}$, $\ket{2}$ and $\ket{3}$ are the slow levels. The goal is to eliminate the fast level and derive an effective Hamiltonian for the slow levels.\\
The first term represents the effective Hamiltonian for the slow levels, while the second term accounts for additional time-dependent terms arising from the unitary transformation \cite{Adhikary2021}.\\
To eliminate the fast level, we make an adiabatic approximation by assuming that the fast level instantaneously adjusts to their eigenstates for each value of the slow variables. This assumption implies that the fast level is always in their instantaneous eigenstates as dictated by the instantaneous values of the slow variables.\\
Under this adiabatic approximation, we neglect the time derivatives of the fast level in the transformed Hamiltonian. This simplification effectively eliminates the fast levels, leading to an effective Hamiltonian solely involving the slow levels \cite{Brion2007}.\\
After neglecting the fast level dynamics, the effective Hamiltonian for the slow levels can be obtained by substituting the adiabatically eliminated expressions for the fast levels in the transformed Hamiltonian.\\
It is important to note that the specific form of the effective Hamiltonian will depend on the details of the system and the specific interactions involved. The adiabatic elimination technique provides a method to simplify the dynamics of a multi-level system by effectively eliminating fast variables, thus allowing for a reduced description of the slow variables in terms of an effective Hamiltonian. The adiabatic approximation simplifies the dynamics of a system with fast and slow variables. However, its validity relies on certain conditions being satisfied. Here are the key conditions for the adiabatic approximation:
\begin{enumerate}
    \item \textbf{Timescale Separation:} The most fundamental condition for the adiabatic approximation is a significant separation of timescales between the fast and slow variables. This implies that the dynamics of the fast variables evolve much more rapidly compared to the timescale of variation of the slow variables. The timescale separation ensures that the fast variables can instantaneously adjust to their eigenstates for each value of the slow variables.
    \item \textbf{Adiabaticity Parameter:} The adiabaticity parameter quantifies the rate of change of the slow variables with respect to the timescale of the fast variables. It is given by:
\begin{align*}
\epsilon = \left|\frac{\dot{\omega}}{\omega^2}\right|,
\end{align*}
where $\omega$ is the characteristic frequency of the fast variables and $\dot{\omega}$ represents its time derivative. The adiabatic approximation is valid when the adiabaticity parameter is small $(\epsilon \ll 1)$, indicating that the slow variables change slowly compared to the fast variables.
    \item \textbf{Avoidance of Level Crossings:} The adiabatic approximation assumes that there are no level crossings or avoided crossings between the energy levels associated with the fast variables. Level crossings can lead to non-adiabatic transitions, where the system rapidly switches between different eigenstates, invalidating the adiabatic approximation. Therefore, it is important to ensure that the system parameters and interactions are chosen such that level crossings are avoided or can be neglected \cite{Casabone2015}.
    \item \textbf{Weak Coupling:} The adiabatic approximation typically assumes that the coupling between the fast and slow variables is weak. Strong coupling between the fast and slow variables can induce rapid transitions and mixing between the eigenstates, violating the adiabatic assumption. Therefore, it is often necessary to ensure that the coupling strengths are sufficiently small to maintain the validity of the adiabatic approximation.
\end{enumerate}
It is important to note that violating any of these conditions can lead to breakdown of the adiabatic approximation and may require alternative techniques to describe the system dynamics accurately.\\
The adiabatic approximation is a powerful tool for simplifying the dynamics of complex systems with separable timescales. However, it is crucial to carefully assess the validity of the adiabatic approximation by considering the specific system characteristics and the fulfillment of the conditions mentioned above \cite{Pellizzari1994}.\\

For the detailed derivation of adiabatic Elimination of $|d\rangle$, we start from the four-level Hamiltonian
\begin{equation}
\tilde H =
\sum_{\alpha\in\{a,b,c,d\}}
\hbar\Delta_\alpha\,|\alpha\rangle\langle\alpha|
+
\sum_{i=1,2}
\hbar\delta_i\,a_i^\dagger a_i
+
\hat V_+
+
\hat V_-
+
\hat H_{\rm drive}
\end{equation}

with the perturbative couplings to $|d\rangle$ separated as
\begin{equation}
\hat V_+
=
\hbar g_1\, a_1\,|d\rangle\langle a|
+
\tfrac{\hbar\Omega}{2}\,|d\rangle\langle b|
+
\hbar g_2\, a_2\,|d\rangle\langle c|,
\qquad
\hat V_- = \hat V_+^\dagger
\end{equation}

and the cavity drive
\begin{equation}
\hat H_{\rm drive}
=
\tfrac{\hbar}{2}\Omega_1(a_1+a_1^\dagger)
+
\tfrac{\hbar}{2}\Omega_2(a_2+a_2^\dagger).
\end{equation}

The decay $|d\rangle\to|\alpha\rangle$ at rate $\Gamma_{d\alpha}$ gives jump operators
\begin{equation}
\hat L_{d\alpha}
=
\sqrt{\Gamma_{d\alpha}}\,|\alpha\rangle\langle d|,
\qquad
\alpha=a,b,c,
\qquad
\sum_\alpha\Gamma_{d\alpha}=\Gamma_d.
\end{equation}

\hrulefill

\section*{Step 1 — Projectors and partition}

Define
\begin{equation}
\hat P
=
|a\rangle\langle a|
+
|b\rangle\langle b|
+
|c\rangle\langle c|,
\qquad
\hat Q = |d\rangle\langle d|,
\qquad
\hat P+\hat Q=\mathbf 1.
\end{equation}

Then
\begin{equation}
\hat H_g
\equiv
\hat P\tilde H\hat P
=
\sum_{\alpha=a,b,c}
\hbar\Delta_\alpha|\alpha\rangle\langle\alpha|
+
\hbar\delta_1 a_1^\dagger a_1
+
\hbar\delta_2 a_2^\dagger a_2,
\end{equation}

\begin{equation}
\hat H_e
\equiv
\hat Q\tilde H\hat Q
=
\hbar\Delta_d\,|d\rangle\langle d|
+
\hbar\delta_1 a_1^\dagger a_1
+
\hbar\delta_2 a_2^\dagger a_2,
\end{equation}

\begin{equation}
\hat V_+ = \hat Q\tilde H\hat P,
\qquad
\hat V_-=\hat P\tilde H\hat Q.
\end{equation}

The non-Hermitian Hamiltonian in the excited-state subspace is given as~\cite{PhysRevA.85.032111}
\begin{equation}
\hat H_{\rm NH}
=
\hat H_e
-
\frac{i}{2}
\sum_{\alpha}
\hat L_{d\alpha}^\dagger \hat L_{d\alpha}
=
\hat H_e
-
\frac{i\hbar\Gamma_d}{2}|d\rangle\langle d|,
\end{equation}

since
\begin{equation}
\sum_\alpha
|d\rangle\langle\alpha|\,|\alpha\rangle\langle d|
=
|d\rangle\langle d|.
\end{equation}

\section*{Step 2 — Schrödinger equation in block form}

Writing
\begin{equation}
|\psi\rangle=|\psi_g\rangle+|\psi_e\rangle
\end{equation}

with
\begin{equation}
|\psi_g\rangle=\hat P|\psi\rangle,
\qquad
|\psi_e\rangle=\hat Q|\psi\rangle,
\end{equation}

the eigenvalue problem
\begin{equation}
\tilde H|\psi\rangle=E|\psi\rangle
\end{equation}

becomes
\begin{equation}
\hat H_g|\psi_g\rangle + \hat V_-|\psi_e\rangle
=
E|\psi_g\rangle,
\end{equation}

\begin{equation}
\hat V_+|\psi_g\rangle + \hat H_e|\psi_e\rangle
=
E|\psi_e\rangle.
\end{equation}

Replacing $\hat H_e\to\hat H_{\rm NH}$ to include decay and solving the second line:
\begin{equation}
|\psi_e\rangle
=
(E-\hat H_{\rm NH})^{-1}\hat V_+|\psi_g\rangle.
\end{equation}

Substituting into the first line gives the energy-dependent effective Hamiltonian acting on the ground manifold:
\begin{equation}
\boxed{
\hat H_{\rm eff}(E)\,|\psi_g\rangle
=
\Bigl[
\hat H_g
+
\hat V_-\,(E-\hat H_{\rm NH})^{-1}\,\hat V_+
\Bigr]
|\psi_g\rangle
=
E|\psi_g\rangle.
}
\end{equation}

Following the effective-operator formalism of Reiter--S{\o}rensen~\cite{PhysRevA.85.032111}, we evaluate the resolvent $(E-\hat H_{\rm NH})^{-1}$ separately for each ground-state eigenstate of $\hat H_g$.
\section*{Step 3 — Diagonalizing $\hat H_g$ and applying state-dependent propagators}

$\hat H_g$ is already diagonal in the basis $\{|\alpha,n_1,n_2\rangle\}$ with eigenvalues
\begin{equation}
E_{\alpha,n_1,n_2}
=
\hbar\bigl(
\Delta_\alpha + n_1\delta_1 + n_2\delta_2
\bigr).
\end{equation}

For each label $l=(\alpha,n_1,n_2)$ define the projector
\begin{equation}
\hat P_l
=
|\alpha,n_1,n_2\rangle
\langle\alpha,n_1,n_2|
\end{equation}

and the channel-restricted excitation
\begin{equation}
\hat V_+^{(l)}
\equiv
\hat V_+ \hat P_l.
\end{equation}
Following Reiter--S{\o}rensen~\cite{PhysRevA.85.032111}, we introduce the state-dependent propagator
\begin{equation}
\bigl[\hat H_{\rm NH}^{(l)}\bigr]^{-1}
\equiv
\bigl(\hat H_{\rm NH} - E_l\bigr)^{-1}.
\end{equation}

The symmetrized (Hermitian) effective Hamiltonian is then:
\begin{equation}
\hat H_{\rm eff}
=
-\frac{1}{2}
\Bigl[
\hat V_-
\sum_l
\bigl(
\hat H_{\rm NH}-E_l
\bigr)^{-1}
\hat V_+^{(l)}
+
{\rm H.c.}
\Bigr]
+
\hat H_g.
\end{equation}

\section*{Step 4 — Action of the propagator}

Compute $\hat V_+^{(l)}$ for $l=(a,n_1,n_2)$:
\begin{align}
\hat V_+|\alpha,n_1,n_2\rangle
&=
\hbar g_1\sqrt{n_1}\,
|d,n_1{-}1,n_2\rangle\,\delta_{\alpha,a}
\nonumber\\
&\quad
+
\tfrac{\hbar\Omega}{2}\,
|d,n_1,n_2\rangle\,\delta_{\alpha,b}
\nonumber\\
&\quad
+
\hbar g_2\sqrt{n_2}\,
|d,n_1,n_2{-}1\rangle\,\delta_{\alpha,c}.
\end{align}

Acting on $|d,m_1,m_2\rangle$ gives
\begin{equation}
\hat H_{\rm NH}|d,m_1,m_2\rangle
=
\hbar\bigl(
\Delta_d + m_1\delta_1 + m_2\delta_2 - i\Gamma_d/2
\bigr)
|d,m_1,m_2\rangle.
\end{equation}

Therefore
\begin{align}
&
\bigl(
\hat H_{\rm NH}-E_l
\bigr)
|d,n_1{-}1,n_2\rangle
\nonumber\\
&=
\hbar
\Bigl[
\Delta_d
+
(n_1{-}1)\delta_1
+
n_2\delta_2
-
\tfrac{i\Gamma_d}{2}
-
\Delta_a
-
n_1\delta_1
-
n_2\delta_2
\Bigr]
|d,n_1{-}1,n_2\rangle.
\end{align}

The photon-number contributions cancel:
\begin{equation}
\bigl(
\hat H_{\rm NH}-E_l
\bigr)
|d,n_1{-}1,n_2\rangle
=
\hbar
\bigl(
\Delta_d - \Delta_a - \delta_1 - \tfrac{i\Gamma_d}{2}
\bigr)
|d,n_1{-}1,n_2\rangle.
\end{equation}

The same analysis for all channels gives
\begin{equation}
\boxed{
\bigl(
\hat H_{\rm NH}-E_l
\bigr)^{-1}
\hat V_+^{(\alpha)}
|\alpha,n_1,n_2\rangle
=
\frac{1}{\hbar\,D_\alpha}
\hat V_+^{(\alpha)}
|\alpha,n_1,n_2\rangle
}
\end{equation}

where the complex detunings are
\begin{equation}
D_a
=
\Delta_d - \Delta_a - \delta_1 - \tfrac{i\Gamma_d}{2},
\end{equation}

\begin{equation}
D_b
=
\Delta_d - \Delta_b - \tfrac{i\Gamma_d}{2},
\end{equation}

\begin{equation}
D_c
=
\Delta_d - \Delta_c - \delta_2 - \tfrac{i\Gamma_d}{2}.
\end{equation}

Thus
\begin{equation}
\hat V_-
\sum_l
(\hat H_{\rm NH}-E_l)^{-1}
\hat V_+^{(l)}
=
\sum_{\alpha,\beta\in\{a,b,c\}}
\frac{1}{\hbar\,D_\alpha}
\hat V_-^{(\beta)}
\hat V_+^{(\alpha)}.
\end{equation}

\section*{Step 5 — Hermitization}

Adding the Hermitian conjugate:
\begin{align}
&
-\frac{1}{2}
\Bigl[
\hat V_-
\sum_l
(\hat H_{\rm NH}-E_l)^{-1}
\hat V_+^{(l)}
+
{\rm H.c.}
\Bigr]
\nonumber\\
&=
-\frac{1}{2}
\sum_{\alpha,\beta}
\hat V_-^{(\beta)}
\hat V_+^{(\alpha)}
\Bigl[
\frac{1}{\hbar D_\alpha}
+
\frac{1}{\hbar D_\beta^{*}}
\Bigr].
\end{align}

Therefore
\begin{equation}
\boxed{
\hat H_{\rm eff}
=
\hat H_g
-
\frac{1}{2\hbar}
\sum_{\alpha,\beta\in\{a,b,c\}}
\hat V_-^{(\beta)}
\hat V_+^{(\alpha)}
\Bigl[
\frac{1}{D_\alpha}
+
\frac{1}{D_\beta^{*}}
\Bigr].
}
\end{equation}

\section*{Step 6 — Operator products $\hat V_-^{(\beta)}\hat V_+^{(\alpha)}$}

Using
\begin{equation}
|\beta\rangle\langle d|\,|d\rangle\langle\alpha|
=
|\beta\rangle\langle\alpha|,
\end{equation}

we obtain

\begin{equation}
\hat V_-^{(a)}\hat V_+^{(a)}
=
\hbar^2|g_1|^2
a_1^\dagger a_1
|a\rangle\langle a|,
\end{equation}

\begin{equation}
\hat V_-^{(b)}\hat V_+^{(b)}
=
\tfrac{\hbar^2|\Omega|^2}{4}
|b\rangle\langle b|,
\end{equation}

\begin{equation}
\hat V_-^{(c)}\hat V_+^{(c)}
=
\hbar^2|g_2|^2
a_2^\dagger a_2
|c\rangle\langle c|.
\end{equation}

Cross terms:
\begin{equation}
\hat V_-^{(b)}\hat V_+^{(a)}
=
\tfrac{\hbar^2 g_1\Omega^*}{2}
a_1
|b\rangle\langle a|,
\end{equation}

\begin{equation}
\hat V_-^{(a)}\hat V_+^{(b)}
=
\tfrac{\hbar^2 g_1^*\Omega}{2}
a_1^\dagger
|a\rangle\langle b|,
\end{equation}

\begin{equation}
\hat V_-^{(b)}\hat V_+^{(c)}
=
\tfrac{\hbar^2 g_2\Omega^*}{2}
a_2
|b\rangle\langle c|,
\end{equation}

\begin{equation}
\hat V_-^{(c)}\hat V_+^{(b)}
=
\tfrac{\hbar^2 g_2^*\Omega}{2}
a_2^\dagger
|c\rangle\langle b|,
\end{equation}

\begin{equation}
\hat V_-^{(c)}\hat V_+^{(a)}
=
\hbar^2 g_1 g_2^*
a_2^\dagger a_1
|c\rangle\langle a|,
\end{equation}

\begin{equation}
\hat V_-^{(a)}\hat V_+^{(c)}
=
\hbar^2 g_1^* g_2
a_1^\dagger a_2
|a\rangle\langle c|.
\end{equation}

\section*{Step 7 — Diagonal contributions: AC Stark shifts}

For $\alpha=\beta$:
\begin{equation}
\frac{1}{D_\alpha}
+
\frac{1}{D_\alpha^*}
=
2\,{\rm Re}\!\left[
\frac{1}{D_\alpha}
\right].
\end{equation}

State $|a\rangle$:
\begin{align}
\hat H_{\rm eff}^{(a,a)}
&=
-\frac{1}{2\hbar}
\cdot
\hbar^2|g_1|^2
a_1^\dagger a_1
|a\rangle\langle a|
\cdot
2\,{\rm Re}\!\left[
\frac{1}{D_a}
\right]
\nonumber\\
&=
-\hbar|g_1|^2
{\rm Re}\!\left[
\frac{1}{D_a}
\right]
a_1^\dagger a_1
|a\rangle\langle a|.
\end{align}

State $|b\rangle$:
\begin{equation}
\hat H_{\rm eff}^{(b,b)}
=
-\frac{\hbar|\Omega|^2}{4}
{\rm Re}\!\left[
\frac{1}{D_b}
\right]
|b\rangle\langle b|.
\end{equation}

State $|c\rangle$:
\begin{equation}
\hat H_{\rm eff}^{(c,c)}
=
-\hbar|g_2|^2
{\rm Re}\!\left[
\frac{1}{D_c}
\right]
a_2^\dagger a_2
|c\rangle\langle c|.
\end{equation}

\section*{Step 8 — Off-diagonal contributions: Raman couplings}

For the $a\leftrightarrow b$ channel:
\begin{align}
\hat H_{\rm eff}^{(a,b)}
&=
-\frac{\hbar\,g_1\Omega^*}{4}
\left[
\frac{1}{D_a}
+
\frac{1}{D_b^{*}}
\right]
a_1|b\rangle\langle a|
+
{\rm H.c.}
\end{align}

Similarly,
\begin{equation}
\hat H_{\rm eff}^{(c,b)}
=
-\frac{\hbar\,g_2\Omega^*}{4}
\left[
\frac{1}{D_c}
+
\frac{1}{D_b^{*}}
\right]
a_2|b\rangle\langle c|
+
{\rm H.c.}
\end{equation}

and
\begin{equation}
\hat H_{\rm eff}^{(a,c)}
=
-\frac{\hbar\,g_1 g_2^*}{2}
\left[
\frac{1}{D_a}
+
\frac{1}{D_c^{*}}
\right]
a_2^\dagger a_1
|c\rangle\langle a|
+
{\rm H.c.}
\end{equation}

\section*{Step 9 — Total effective Hamiltonian}

Combining all terms:
\begin{align}
\hat H_{\rm eff}
&=
\hat H_g
+
\hat H_{\rm drive}
\nonumber\\
&\quad
-
\hbar|g_1|^2
{\rm Re}\!\left[
\tfrac{1}{D_a}
\right]
a_1^\dagger a_1
|a\rangle\langle a|
\nonumber\\
&\quad
-
\tfrac{\hbar|\Omega|^2}{4}
{\rm Re}\!\left[
\tfrac{1}{D_b}
\right]
|b\rangle\langle b|
\nonumber\\
&\quad
-
\hbar|g_2|^2
{\rm Re}\!\left[
\tfrac{1}{D_c}
\right]
a_2^\dagger a_2
|c\rangle\langle c|
\nonumber\\
&\quad
-
\tfrac{\hbar g_1\Omega^*}{4}
\left[
\tfrac{1}{D_a}
+
\tfrac{1}{D_b^{*}}
\right]
a_1|b\rangle\langle a|
\nonumber\\
&\quad
-
\tfrac{\hbar g_2\Omega^*}{4}
\left[
\tfrac{1}{D_c}
+
\tfrac{1}{D_b^{*}}
\right]
a_2|b\rangle\langle c|
\nonumber\\
&\quad
-
\tfrac{\hbar g_1 g_2^*}{2}
\left[
\tfrac{1}{D_a}
+
\tfrac{1}{D_c^{*}}
\right]
a_2^\dagger a_1
|c\rangle\langle a|
+
{\rm H.c.}
\end{align}

\section*{Step 10 — Effective Lindblad operators}

For each spontaneous-decay channel:
\begin{equation}
\hat L_{\rm eff,d\alpha}
=
\hat L_{d\alpha}
\sum_l
\bigl(
\hat H_{\rm NH}-E_l
\bigr)^{-1}
\hat V_+^{(l)}.
\end{equation}

Substituting:
\begin{align}
\hat L_{\rm eff,d\alpha}
&=
\sqrt{\Gamma_{d\alpha}}
\,|\alpha\rangle\langle d|
\sum_\mu
\frac{\hat V_+^{(\mu)}}{\hbar D_\mu}
\nonumber\\
&=
\sqrt{\Gamma_{d\alpha}}
\Bigl[
\frac{g_1 a_1}{D_a}
|\alpha\rangle\langle a|
+
\frac{\Omega/2}{D_b}
|\alpha\rangle\langle b|
+
\frac{g_2 a_2}{D_c}
|\alpha\rangle\langle c|
\Bigr].
\end{align}

Thus
\begin{equation}
\boxed{
\hat L_{\rm eff,d\alpha}
=
\sqrt{\Gamma_{d\alpha}}
\Bigl[
\frac{g_1 a_1}{D_a}
|\alpha\rangle\langle a|
+
\frac{\Omega/2}{D_b}
|\alpha\rangle\langle b|
+
\frac{g_2 a_2}{D_c}
|\alpha\rangle\langle c|
\Bigr].
}
\end{equation}

The remaining dissipators
\begin{equation}
\hat L_{\kappa_i}
=
\sqrt{\kappa_i}\,a_i,
\qquad
\hat L_{\gamma_1}
=
\sqrt{\gamma_1}\,|b\rangle\langle a|,
\qquad
\hat L_{\gamma_2}
=
\sqrt{\gamma_2}\,|b\rangle\langle c|
\end{equation}

already act entirely in the ground manifold and remain unchanged.

\bibliographystyle{apsrev4-2}
\bibliography{Biblio}

\begin{thebibliography}{99}
\section{References}
\bibitem{1}
Antonio de Freitas, L. Sanz and José M. Villas-Bôas (2017): Coherent control of the dynamics of a single quantum-dot exciton qubit in a cavity. Phys. Rev. B 95, 115110.
\bibitem{2}
Arif Warsi Laskar et. al, 2020 https://doi.org/10.1103/PhysRevLett.125.013601
\bibitem{3}
Christopher C. Gerry and J. H. Eberly:
Phys. Rev. A 42, 6805 – Published 1 December 1990
\bibitem{4}
Holmes, C. A., C. P. Meaney and G. J. Milburn (2012): Synchronization of many
nanomechanical resonators coupled via a common cavity field. Phys. Rev. E 85,
066203.
\bibitem{5}
Lee, T. E. and M. C. Cross (2013): Quantum-classical transition of correlations of
two coupled cavities. Phys. Rev. A 88, 013834.
\bibitem{6}
Pikovsky, A., M. Rosenblum and J. Kurths (2003): Synchronization. A universal
concept in nonlinear sciences (Cambridge University Press, Cambridge).
\bibitem{7}
Roulet, A. and C. Bruder (2018a): Synchronizing the Smallest Possible System. Phys.
Rev. Lett. 121, 053601.
\bibitem{8}
Roulet, A. and C. Bruder (2018b): Quantum Synchronization and Entanglement Generation. Phys. Rev. Lett. 121, 063601
\bibitem{9}
van der Pol, B. (1926): On relaxation oscillators. Phil. Mag. 2, 978.
\bibitem{10}
van der Pol, B. and J. van der Mark (1928): The heartbeat considered as a relaxation
oscillation and an electrical model of the heart. Phil. Mag. 6, 763.
\bibitem{11}
Walter, S., A. Nunnenkamp and C. Bruder (2014): Quantum Synchronization of a
Driven Self-Sustained Oscillator. Phys. Rev. Lett. 112, 094102.
\bibitem{12}
Zhirov, O. V. and D. L. Shepelyansky (2006): Quantum synchronization. E. Phys. J.
D 38, 375.
\bibitem{13}
https://journals.aps.org/pra/pdf/10.1103/PhysRevA.12.1019
\bibitem{PhysRevA.42.6805}
Gerry CC, Eberly JH. Dynamics of a Raman coupled model interacting with two quantized cavity fields. Phys Rev A. 1990 Dec 1;42(11):6805-6815. doi: 10.1103/physreva.42.6805. PMID: 9903981.
\bibitem{PhysRevA.103.033310}
PhysRevA.103.033310
\bibitem{PhysRevLett.107.023601}
PhysRevLett.107.023601
\bibitem{PhysRevLett.114.023602}
PhysRevLett.114.023602
\bibitem{doi:10.1080/09500349414550601}
doi:10.1080/09500349414550601
\end{thebibliography}

\end{document}